\documentclass[twocolumn,aps,amssymb,noshowpacs,superscriptaddress,nofootinbib]{revtex4}

\usepackage{amsmath,amssymb}
\usepackage{graphicx}
\usepackage{verbatim}
\usepackage{amsfonts}
\usepackage{dcolumn}
\usepackage{bm}
\usepackage[section]{placeins} 
\usepackage{color}
\usepackage{mathrsfs}
\usepackage[colorlinks,bookmarks=true,citecolor=blue,linkcolor=red,urlcolor=blue]{hyperref}
\usepackage{afterpage}
\usepackage{ulem}

\usepackage{longtable}
\usepackage{multirow}

\newcommand{\beq}{\begin{eqnarray} }
\newcommand{\eeq}{\end{eqnarray} }
\newcommand{\Beq}{\begin{eqnarray*} }
\newcommand{\Eeq}{\end{eqnarray*} }

\newcommand{\RNum}[1]{\uppercase\expandafter{\romannumeral #1\relax}}

\usepackage{ulem}

\begin{document}
\draft

\title{Unlocking of time reversal, space-time inversion and rotation invariants in magnetic materials}
\author{Jian Yang}
\affiliation{Beijing National Laboratory for Condensed Matter Physics and Institute of Physics, Chinese Academy of Sciences, Beijing 100190, China}
\author{Zheng-Xin Liu}
\email{liuzxphys@ruc.edu.cn}
\affiliation{Department of Physics, Renmin University, Beijing, China}
\author{Chen Fang}
\email{cfang@iphy.ac.cn}
\affiliation{Beijing National Laboratory for Condensed Matter Physics and Institute of Physics, Chinese Academy of Sciences, Beijing 100190, China}

\date{\today}
\begin{abstract}
Time reversal ($T$) and space inversion are symmetries of our universe in the low-energy limit.
Fundamental theorems relate their corresponding quantum numbers to the spin for elementary particles: $\hat{T}^2=(\hat{P}\hat{T})^2=-1$ for half-odd-integer spins and $\hat{T}^2=(\hat{P}\hat{T})^2=+1$ for integer spins.
Here we show that for elementary excitations in magnetic materials, this ``locking'' between quantum numbers is lifted: $\hat{T}^2$ and $(\hat{P}\hat{T})^2$ take all four combinations of $+1$ and $-1$ regardless of the value of the spin, where $T$ now represents the composite symmetry of time reversal and lattice translation.
Unlocked quantum numbers lead to new forms of minimal coupling between these excitations and external fields, enabling novel physical phenomena such as the``cross-Lamor precession'', indirectly observable in a proposed light-absorption experiment.
We list the magnetic space groups with certain high-symmetry momenta where such excitations may be found.
\end{abstract}
\maketitle

\section{Introduction}
Spatial inversion ($P$) and time reversal ($T$) are considered as fundamental symmetries of our universe at low energy (but are broken in weak forces)\cite{Lee1957}.
As operators, inversion acts on the spatial degrees of freedom of particles, and time reversal the internal ones: they naturally commute $[\hat{P},\hat{T}]=0$.
(We use hatted letters for operators in Hilbert space, and unhatted ones for symmetries themselves.)
Restricting the discussion to the single-particle sector of the Hilbert space, we understand that inversion squares to unity, $\hat{P}^2=1$, because (i) two consecutive inversions equal identity and (ii) inversion does not act on spin.
Time reversal, on the other hand, an anti-unitary symmetry reversing time and inverting spin, is represented by $\hat{T}=e^{i\hat{S}_y\pi}K$,
where $\hat{S}_y$ is the $y$-component of the spin operator.
Therefore we have $\hat{T}^2=(-1)^{2s}$, where $s(s+1)\hbar^2=\hat{\mathbf{S}}^2$.

Symbolically, we have
\begin{equation}\label{eq:1}
[\hat{P},\hat{T}]=0, \hat{P}^2=1, \hat{T}^2=(-1)^{2s}.
\end{equation}
Eq.(\ref{eq:1}) relates the space-inversion, the time-reversal and the rotation invariants for particles in vacuum.
These relations can be more concisely represented by three \textit{invariants} $\chi_T:=\hat{T}^2$, $\chi_{PT}:=(\hat{P}\hat{T})^2$ and $\chi_S:=(-1)^{2s}$:
\begin{equation}\label{eq:2}
\chi_T=\chi_{PT}=\chi_S.
\end{equation}
For a given type of particle with spin-$s$, $\chi_T$ and $\chi_{PT}$ are completely fixed to be $+1$ if $s$ is an integer (particle being boson) and $-1$  if $s$ is a half-odd-integer (particle being fermion).
Eq.(\ref{eq:2}) applies for any Lorentz-invariant theory given that $\hat P$ and $\hat T$ do not act on the species degrees of freedom \cite{Streater1989, Weinberg1995}.

In this paper, we show that for elementary excitations in certain lattices with magnetic ordering, the three invariants, $\chi_S$ (the definition of which is modified due to the absence of continuous rotation), $\chi_T$ and $\chi_{PT}$, are independent from each other, and can take all eight combinations, in contrast to being locked with each other as in Eq.(\ref{eq:2}) in real vacuum.
Here, what $T$  represents is not time reversal, broken by the magnetic order, but a composite symmetry of time reversal and some lattice translation, which is present in many antiferromagnets.
The properties of elementary excitations having unlocked invariants are demonstrated through the example of $\mathbf{R}=(\pi/a,\pi/a,\pi/a$) in magnetic space group 222.103, where $a$ is the lattice constant of the magnetic unit cell.
The unlocked invariants enable unconventional linear responses to electromagnetic field.
We show that polarization operators and magnetization operators together furnish an $SO(4)$ algebra unseen in previous studies to our best knowledge.
The $SO(4)$ algebra leads to a new type of Larmor precession, where a precession of polarization is driven by a magnetic field, and a precession of magnetization by an electric field.
We propose a light-absorption experiment to observe this ``cross-Larmor precession''.

\section{Invariants on a non-magnetic lattice}
When we put any theory on a lattice, the three-dimensional-rotation symmetry reduces to a point group symmetry, so the rotation invariant $\chi_S$ is no longer well-defined.
For a modified version, we require the point group in question to contain at least three orthogonal twofold rotation axes.
For example, the group $D_{2h}$ having three such twofold axes $C_{2x,2y,2z}$, satisfies our constraint, while the group $C_{2h}$, having only one, does not.
The redefined rotation invariant $\chi_S$ is
\begin{equation}\label{eq:0}
\hat{C}_{2m}\hat{C}_{2n}=\chi_S\hat{C}_{2n}\hat{C}_{2m},
\end{equation}
where $m\neq{n}$ take values $x,y,z$.

\begin{figure}
\begin{centering}
\includegraphics[width=1\linewidth]{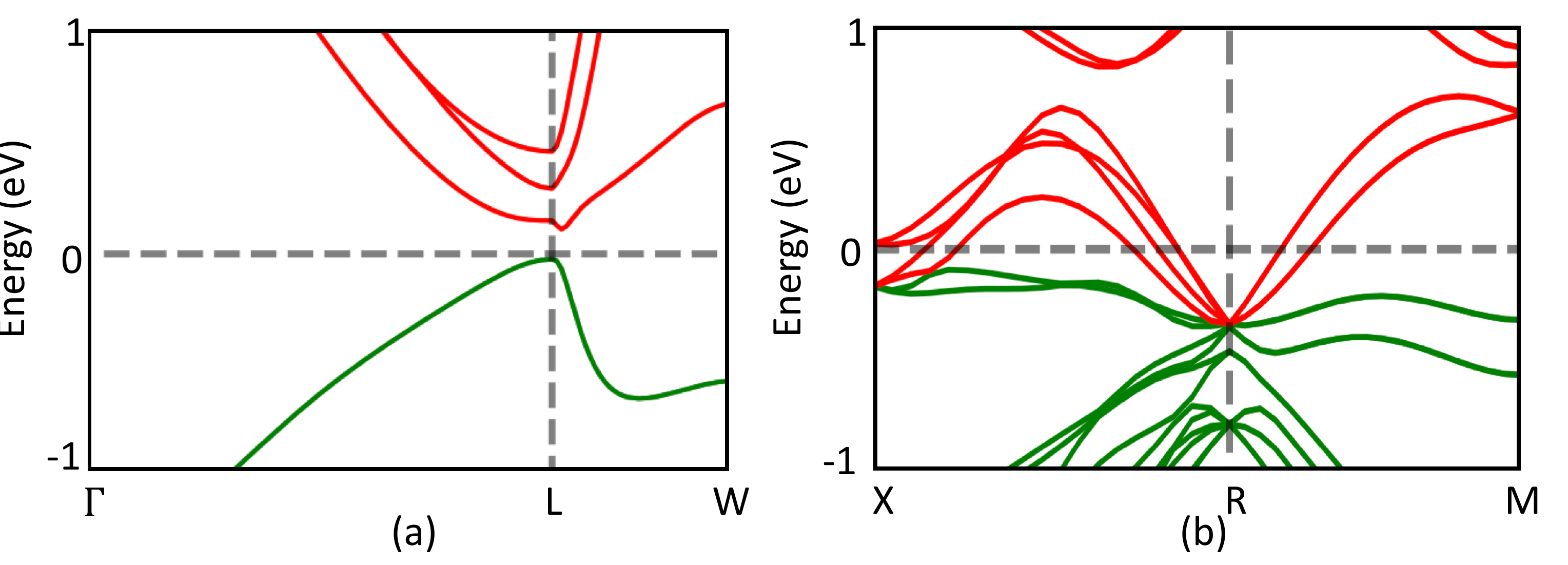}
\par\end{centering}
\protect\caption{\label{fig:1}Electron excitations centering at BZ corners. (a) In SnTe, the conduction bottom (and valence top) is located at L instead of $\Gamma$. (b) In BaPPt, the conduction and the valence bands touch each other at R. In both cases, there are bands of elementary excitations centering at some nonzero and high-symmetry momentum. These band structures are downloaded from http://materiae.iphy.ac.cn}
\end{figure}

In real vacuum, for any stable theory, the lowest particle excitations are near zero momentum, because this is the only special point having higher symmetry than its neighborhood.
However, on a lattice, the elementary excitation may also appear at the {corners of the Brillouin zone} (BZ), which are nonzero crystal momenta with highest symmetry in the neighborhood\cite{Vergniory2017}.
This is the case when, for example, the band bottom of the conduction band appears at a BZ corner [Fig.\ref{fig:1}(a)], or when the conduction and the valence bands touch each other at some BZ corner [Fig.\ref{fig:1}(b)].

Yet another complication in lattices is that some spatial symmetries are neither point-group operations nor lattice translations, but are composite operations of a point group symmetry and a fractional lattice translation.
Lattice symmetry groups containing these spatial symmetries are called nonsymmorphic space groups\cite{Bradley2010}.
Consider for example space group $Pbca$, generated by three orthogonal lattice translations $t_{\mathbf{x},\mathbf{y},\mathbf{z}}$ along $x,y,z$-directions with lattice constants $a,b,c$, respectively, space inversion about the origin, $P$, and three twofold screw axes parallel to $x,y,z$-axes, $C_{2x}\to 2_{100}t_{\mathbf{x}/2+\mathbf{y}/2}, C_{2y}\to 2_{010}t_{\mathbf{y}/2+\mathbf{z}/2}, C_{2z}\to 2_{001}t_{\mathbf{x}/2+\mathbf{z}/2}$; $2_{mnl}$ represents a $\pi$-rotation on both spatial coordinates and spin components about the $[mnl]$-direction, where $[mnl]$ are Miller indices.
$t_{d_1\mathbf{x}+d_2\mathbf{y}+d_3\mathbf{z}}$ represents translation by $d_1\mathbf{x}a+d_2\mathbf{y}b+d_3\mathbf{z}c$ where $\mathbf{x}, \mathbf{y}, \mathbf{z}$ are unit vectors and $a, b, c$ lattice constants.
It is straightforward to check that
$\hat C_{2x}\hat C_{2y}=(-1)^{2s_0}\hat C_{2y}\hat C_{2x}\hat t_{-\mathbf{x}+\mathbf{y}+\mathbf{z}}$.
Here $s_0$ is the physical spin of the particle, $s_0\in{even}$ for bosons and $s_0\in{odd}$ for fermions.
Consider excitations at $\mathbf R=(\pi/a,\pi/b,\pi/c)$, where $\hat{t}_{-\mathbf{x}+\mathbf{y}+\mathbf{z}}=-1$, then we have $\chi_S=-(-1)^{2s_0}$.
The additional minus sign means that the half-lattice translations give a ``twist'' to the rotation invariants, making integer spins look like half-odd-integer spins, and vice versa\cite{Bradlyn2016}.

However, on a nonmagnetic, centrosymmetric lattice, similar twist does not occur in the invariants $(\chi_T,\chi_{PT})$.
For excitations on any nonmagnetic lattice, we always have $\chi_T=\chi_{PT}=(-1)^{2s_0}$.
Therefore, for excitations at $\mathbf R=(\pi/a,\pi/b,\pi/c)$ in space group $Pbca$, we have
\begin{equation}\label{eq:3}
\chi_T=\chi_{PT}=-\chi_S.
\end{equation}

\section{Further unlocking of invariants on a magnetic lattice}
Is it possible to further unlock the values of $\chi_T$ and $\chi_{PT}$ from the constraint $\chi_T=\chi_{PT}$?
We seek this possibility in elementary excitations in magnetically ordered lattices, the band topology of which has recently become a research focus\cite{Mong2010,Fang2013,Fang2014,Watanabe2018,Hua2018,Cano2019,Xu2020,Zou2020}.
While the physical time reversal is broken by magnetism, many antiferromagnetic materials preserve a symmetry in the form $T\rightarrow{T}_{phy}t_{\mathbf{L}/2}$, where $T_{phy}$ is pure time reversal operator acting only on internal degrees of freedom, and $\mathbf{L}/2$ is a lattice vector (which becomes a half-lattice vector in the magnetic unit cell)\cite{Mong2010,Fang2013}.
Symmetries of magnetic systems are classified by magnetic space groups, and when the above $T$ is a symmetry, the magnetic space groups are called type-IV\cite{Bradley1968} ,which are of our interest.
When this is the case, excitations at any time-reversal invariant momentum in the magnetic Brillouin zone (mBZ) have this $T$ symmetry.
We then naturally redefine $\chi_T$ as $\chi_T\equiv\hat{T}^2=(\hat{T}_{phy}\hat{t}_{\mathbf{L}/2})^{2}$.
As expected, the translation part in $T$ in general changes the value of $\chi_T$, and could also make $P$ and $T$ not commute\cite{Mong2010,Cano2019}.
One can use the theory of projective representation(Rep)s to show that $(\chi_T,\chi_{PT})$ can take all possible values of $(+1,+1)$, $(+1,-1)$, $(-1,+1)$ and $(-1,-1)$, irrespective of the rotation invariant $\chi_S$.
Let us use an example to illustrate one of these possibilities here, and defer the general proof in Table \ref{SSGIV} of appendix.

In Fig.\ref{fig:2}, a conjectured magnetic structure of NdZn\cite{Gallego2016,Xu2020} (magnetic space group 222.103) is shown.
Looking from $O$ or $O'$, the magnetic structure is invariant under all proper rotations of a cube (point group $O$); the structure is also centro-symmetric about $I$, and has a composite symmetry $T\rightarrow{T}_{phy}t_{(\mathbf{x}+\mathbf{y}+\mathbf{z})/2}$, where ${(\mathbf{x}+\mathbf{y}+\mathbf{z})/2}$ is a lattice vector of the nonmagnetic lattice.
It is straightforward to check ${\hat{T}}^2=(-1)^{2s_0}\hat{t}_{\mathbf{x}+\mathbf{y}+\mathbf{z}}$, $\hat{P}\hat{T}=\hat t_{\mathbf{x}+\mathbf{y}+\mathbf{z}}\hat{T}\hat{P}$, and $\chi_S=(-1)^{2s_0}$.
Therefore, at mBZ corner $\mathbf R=(\pi/a,\pi/a,\pi/a)$, we have
\begin{equation}
\chi_T=-\chi_{PT}=-\chi_S.
\end{equation}
For magnons on this lattice, we have $(\chi_S, \chi_T,\chi_{PT})=(+1,-1,+1)$, and for electrons $(-1,+1,-1)$, neither of which is possible on nonmagnetic lattices.
\begin{figure}
\begin{centering}
\includegraphics[width=1\linewidth]{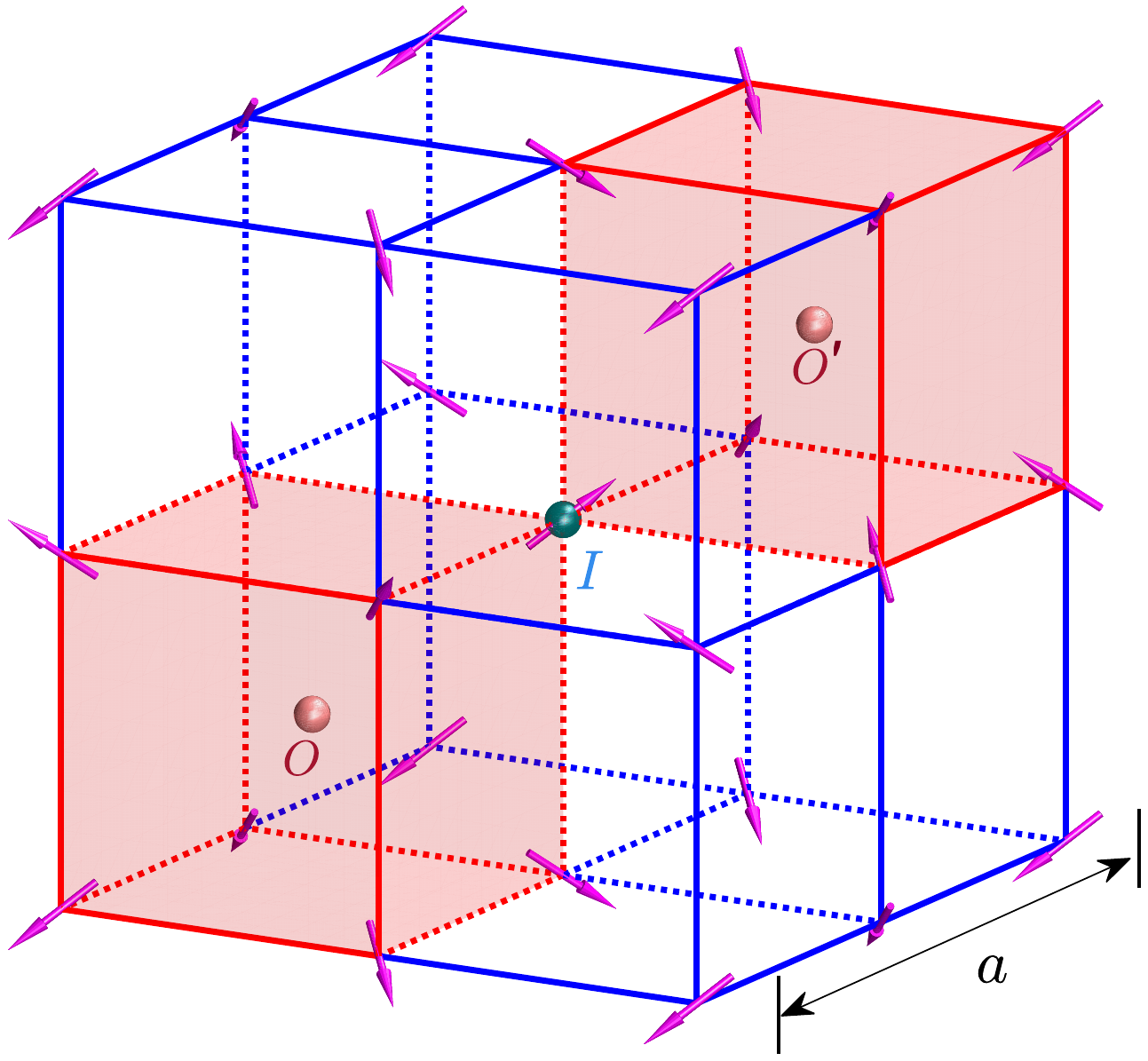}
\par\end{centering}
\protect\caption{\label{fig:2}The magnetic unit cell with lattice constant $a$ of a magnetic structure having magnetic space group 222.103, conjectured as the symmetry for the magnetic ground state of NdZn. Yellow circles $O, O'$ are body centers of point group symmetry $O$, and orange circle $I$ is an inversion center. $T$ is time reversal combined with a lattice translation that relates the two cubes outlined in red. The local moments are polarized so that they are all-out with respect to $O$ and all-in with respect to $O'$.}
\end{figure}

\section{Effective theory with minimal coupling}
In quantum field theory, particles couple to an external source (field) if there exists a bilinear of particle field operators that carries the same symmetry Rep as the one carried by the source.
The form of coupling hence depends on the projective Rep carried by particle field, as well as the linear Rep carried by the external field.
The symmetry invariants $\chi_{S,T,PT}$ do not denote specific Reps, but classes of Reps.
For example, for group $SO(3)$, all integer spins belong to one class ($\chi_S=1$) while all half-odd-integer spins to another class ($\chi_S=-1$).

We postpone the former definition of ``classes of Reps'' and invariants to a later part.
For now, we only need that fact that new invariants unlocked in magnetic systems allow new classes of Reps for field operators, which in turn bring about new forms of linear coupling to external sources such as applied electromagnetic field and strain tensor field.
In this section we restrict the discussion to electron excitations, $s_0=1/2$, linearly coupled to an electromagnetic field.
Physically, since electrons have charge, the electromagnetic field has two effects: the orbital effect that changes the crystal momentum of the electron in a fixed band structure, and the Zeeman effect which modifies the periodic part of Bloch wavefunctions at each momentum, thereby modifying the band structure itself\cite{Anderson1997}.
The invariants, $\chi_{S,T,PT}$, have little to do with the first effect, so we focus on the second, altogether ignoring the change of crystal momentum by the external fields.

Again let us use the magnetic lattice in Fig.\ref{fig:2} as the example, and focus on mBZ corner $\mathbf R=(\pi/a,\pi/a,\pi/a)$ having little co-group $O_h\times Z_{2}^{T}$.
There are three different irreducible Reps, $R_{1/2}$ and $R_{3/2}^\pm$, of $G_\mathbf{R}$, all satisfying $(\chi_S,\chi_{T},\chi_{PT})=(-1,+1,-1)$.
Here we pick one Rep $R_{1/2}$ for our analysis, where the generators are given by
\begin{eqnarray}\label{eq:5}
\hat{T}&=&\tau_y \sigma_yK,\\
\nonumber
\hat{P}&=&i\tau_y,\\
\nonumber
\hat{C}_{2x}&=&i\sigma_x ,\\
\nonumber
\hat{C}_{31}&\equiv&\hat{C}_{3,111}=-\exp(-i\sigma_{111}\pi/3),\\
\nonumber
\hat{C}_{4z}&=&i\tau_z\exp(-i\sigma_z\pi/4),
\end{eqnarray}
where $C_{3,111}$ is the threefold rotation about the $[111]$-direction, and $\sigma_{111}\equiv(\sigma_x+\sigma_y+\sigma_z)/\sqrt{3}$.
Group theory allows us to classify all hermitian bilinears of field operators in the form $ \psi^\dag \tau_\mu\sigma_\nu \psi$ into the Reps of the symmetry group as in Table \ref{Tab:1}.
On the other hand, all monomials of crystal momentum relative to $\mathbf R$, $\mathbf{q}=\mathbf{k}-\mathbf R$, and the components of electric, magnetic and strain fields can also be put into these Reps, also summarized in Table \ref{Tab:1}.

\begin{table*}[htbp]
\caption{Classification of hermitian operators, monomials of relative momentum and external electric, magnetic and strain tensor fields into their respective irreducible Reps of $O_h \times Z_{2}^{T}$, for a minimal effective theory having symmetries given in Eq.(\ref{eq:5}). The symbols of the irreducible Reps follow the conventional definition of Ref.[\onlinecite{Bradley2010}], and the $\pm$ in superscript denotes whether this Rep changes sign under $\hat{T}$. $\mathbf B\star \mathbf q$ stands for $(B_yq_z+B_zq_y, B_zq_x+B_xq_z, B_xq_y+B_yq_x)$.} \label{Tab:1}
\centering
{%\tiny
 \begin{tabular}{ |c|c|c|c|c|c|c|c|c|}
 \hline
 Irreducible Rep&$A_{2u}^{-}$&$A_{2g}^{-}$&$A_{1u}^{-}$&$T_{2u}^{+}$&$T_{2g}^{+}$&$T_{1u}^{+}$&$T_{1g}^{-}$&$A_{1g}^{+}$\\
 \hline
bilinear Operators&$\psi^\dag \tau_{x}\psi$&$\psi^\dag\tau_{y}\psi$&$\psi^\dag\tau_{z}\psi$&$\psi^\dag\tau_{x}\sigma_{i}\psi$&$\psi^\dag\tau_{y}\sigma_{i}\psi$&$\psi^\dag\tau_{z}\sigma_{i}\psi$
&$\psi^\dag\sigma_{i}\psi$&$\psi^\dag\psi$\\
\hline
Applied fields&& $B_x B_y B_z$ &$\mathbf{E}\cdot \mathbf{B}$&
&$(E_{y}E_{z},E_{x}E_{z},E_{x}E_{y})$
&$E_{i}$&$B_{i}$&$E^{2}$\\

&&&&
&$(B_{y}B_{z},B_{x}B_{z},B_{x}B_{y})$
&&&$B^{2}$\\

&&&&
&$(\epsilon_{yz},\epsilon_{xz},\epsilon_{xy})$
&&&\\
\hline
Monomials of momentum& $q_xq_yq_z$ &&&&$(q_{y}q_{z},q_{x}q_{z},q_{x}q_{y})$&&&$q^{2}$\\
\hline
cross coupling &$\epsilon_{yz}q_x+\epsilon_{zx}q_y+\epsilon_{xy}q_z$&&&$(\mathbf B{\star}\mathbf q)_i$&&$(\mathbf B\times \mathbf q)_i$&$(\mathbf E\times \mathbf q)_i$&$\mathbf E\times\mathbf B\cdot\mathbf q$\\
\hline
\end{tabular}
}
\end{table*}

Matching the Reps between fermion bilinear operators and monomials of momenta and of fields, we find the effective theory with minimal couplings.
The free part of the theory is obtained by matching the monomials of momenta to the bilinears:
\begin{equation}\label{eq:free}
h_0[\psi]=-\frac{1}{2m}\int{dr^3}\psi^\dag\tau_y(\sigma_x\partial^2_{yz}+\sigma_y\partial^2_{zx}+\sigma_z\partial^2_{xy})\psi.
\end{equation}
The minimal coupling terms to the electric field are
\begin{equation}\label{eq:6}
h_{E}[\psi,\mathbf{E}]=-\lambda_E\mathbf{E}\cdot \int{dr^3}\psi^\dag\tau_z \pmb{\sigma} \psi;
\end{equation}
the minimal coupling terms to the magnetic field
\begin{equation}\label{eq:7}
h_{B}[\psi,\mathbf{B}]=-\lambda_B\mathbf{B}\cdot\int{dr^3}\psi^\dag \pmb \sigma\psi,
\end{equation}
where $\lambda_{E,B}$ are coupling constants.
From Eq.(\ref{eq:6},\ref{eq:7}), we have the effective polarization and magnetization operators:
\begin{eqnarray}\label{eq:8}
\hat{P}_i&=&-\frac{\delta\left({h_0+h_E+h_B}\right)}{\delta{E}_i}=\lambda_E\tau_z\sigma_i,\\
\nonumber
\hat{M}_i&=&-\frac{\delta\left({h_0+h_E+h_B}\right)}{\delta{B}_i}=\lambda_B\sigma_i.
\end{eqnarray}
We note that Eq. (\ref{eq:8}) implies the quantum nature of polarization $\hat{P}_i$ in this minimal effective theory, as different components do not commute
\begin{equation}\label{eq:9}
[\hat{p}_i,\hat{p}_j]=i\epsilon_{ijk}\hat{m}_k, [\hat{p}_i,\hat{m}_j]=i\epsilon_{ijk}\hat{p}_k,
\end{equation}
where we have defined the dimensionless reduced dipole operator $\hat{p}_i\equiv\hat{P}_i/(2\lambda_E)$ and magnetization operator $\hat{m}_i\equiv\hat{M}_i/(2\lambda_B)$.

\section{Hidden $SO(4)$ algebra, cross-Larmor precession and absorption}
The commutation relations in Eq.(\ref{eq:9}) imply that $\hat{p}_i$ and $\hat{m}_i$ together form an $SO(4)$ algebra.
Consider one quasiparticle magnetically polarized in the $z$-direction, that is, $\hat{m}_z|\psi(q=0)\rangle=|\psi(q=0)\rangle$.
At $t=0$, an electric field was applied along $x$-direction.
According to Eq.(\ref{eq:9}), at $t>0$, we have
\begin{equation}\label{eq:psit}
[\cos(2\lambda_Et)\hat{m}_z+\sin(2\lambda_Et)\hat{p}_y]|\psi(t)\rangle=|\psi(t)\rangle.
\end{equation}
Specially, at $t=(n-1/4)\pi/\lambda_E$, the quasiparticle is completely electrically polarized along $y$-direction, and is spin unpolarized $\langle\psi(t)|\hat{m}_i|\psi(t)\rangle=0$.
Therefore an electric field rotates a magnetic dipole into an electric dipole, hence the name ``cross-Larmor precession''.

How could cross-Larmor precession be observed in experiments?
The original Larmor precession is observed in the absorption in ferrogmagnets\cite{Griffiths1946}, where two elements are essential: a polarized spin configuration and an oscillating magnetic field component perpendicular to the magnetization.
In our system, we first use a magnetic field $\mathbf{B}_0$ to spin-polarize the electrons at $\mathbf{q}=0$ along $z$-direction.
Then we use a polarized light propagating along $y$-direction, the $\mathbf{E}$-vector of which is polarized in the $x$-direction.
With this geometry the Larmor precession is minimized, because the $\mathbf{B}$-vector of the light is parallel to the spin polarization.
The cross-Larmor precession, on the other hand, is maximized because the electric field is perpendicular to the spin polarization.
We therefore predict a resonant absorption edge at a frequency $|2\lambda_B\mathbf{B}_0|$, assuming that the system be exactly at half-filling.
This is what we call a cross-resonant absorption, because the level splitting is obtained using a magnetic field, while the resonance is realized using an electric field.
While $SO(4)$ algebra in Eq.(\ref{eq:9}) is a sufficient, but not necessary, condition for the presence of cross-resonant absorption.
An example illustrating the non-necessity is shown in sec.\ref{3/2reps} of appendix, where the other two Reps at $\mathbf R$ in 222.103 are analyzed.
There we show, though the $SO(4)$ algebra no longer holds, the cross-resonant absorption is still present.

We emphasize that neither the $SO(4)$ algebra nor the cross-resonant absorption is restricted to a specific momentum [$\mathbf R=(\pi/a,\pi/a,\pi/a)$] or a specific magnetic space group (222.103).
The little group at $\mathbf{R}$ in 222.103 is $G_\mathbf{R}=O_h\times{Z}_{2}^{T}$.
As one breaks $O_h$ down to $T_h$, $D_{4h}$ and $D_{2h}$, one finds that $R_{1/2}$ remains an irreducible Rep.
When $G_\mathbf{R}=T_h\times{Z}_{2}^{T}$, both the $SO(4)$ algebra and the cross-resonant absorption hold;
and as $G_\mathbf{R}$ is reduced to $D_{4h,2h}\times{Z}_{2}^{T}$, the polarization operators $\hat{p}_{x,y,z}$ and magnetization operators $\hat{m}_{x,y,z}$ no longer form the six generators of $SO(4)$, but the cross-resonant absorption still appears
(See sec.\ref{sysredu} of appendix for the derivation of the above statements).
Due to the preservation of cross-resonant absorption in symmetry reduction, we predict that at least four magnetic space groups (222.103, 223.109, 200.17 and 201.21) in which certain irreducible Reps at $\mathbf{R}$ shows $SO(4)$, and 12 groups (222.103, 223.109, 200.17, 201.21, 126.386, 130.432, 131.446, 135.492, 47.256, 48.264, 55.361 and 56.373) that potentially host cross-resonant absorption.

\section{Guidelines for finding materials}
The four combinations $(\chi_{S},\chi_T,\chi_{PT})=(-1,+1,-1), (+1,-1,+1), (+1,+1,-1), (-1,-1,+1)$ are restricted to magnetic materials, realized as elementary excitations with momentum at corners of the mBZ, and they bring about novel physical effects as are discussed above.
Hence it is vital that we provide clues as to where these excitations may be found in real materials.
In mathematics, invariants such as $\chi_T$, $\chi_{PT}$ and $\chi_{S}$ exactly correspond to {the invariants of the second group cohomology}\cite{Hatcher2005}, $\mathcal{H}^2[G_{\mathbf Q},U(1)]$, of the little co-group $G_{\mathbf{Q}}$, which classify the {projective Reps} of $G_{\mathbf{Q}}$.
Here we insert a few technical comments on $G_{\mathbf{Q}}$.
As we focus on the elementary excitations centered at high-symmetry momentum $\mathbf{Q}$, the full magnetic space group $M$ is reduced to the little group $M(\mathbf{Q})\subset{M}$ that leaves $\mathbf{Q}$ invariant, up to a reciprocal lattice vector.
This little group has lattice translation $Tr$ as its normal subgroup, so that a quotient group can be defined $M(\mathbf{Q})/Tr$, which is always isomorphic to a magnetic point group $G_{\mathbf{Q}}$, called the little co-group of $M$ at $\mathbf{Q}$.
As far as type-IV magnetic space groups are concerned, $G_{\mathbf{Q}}$ has the simple structure of a point group direct-product time reversal.
The invariants of each of the 32 $G_{\mathbf{Q}}$'s can be calculated, and are listed in Table \ref{H2cohomology}.

However, it is not obvious that all the possible values of these invariants may be taken in elementary excitations of real materials.
While we are unable to address this general question, a full answer can be obtained, as discussion is restricted to excitations that form bands (magnon or electron).
Throughout this article, symbols such as $C_{2x}$ or $T$ all refer to elements of some $G_{\mathbf{Q}}$.
Yet, one should keep in mind that physically, $g\in{G}_{\mathbf{Q}}$ represent a coset in $M(\mathbf{Q})/Tr$, out of which a representative element $\tilde{g}\in M(\mathbf{Q})$ may be chosen.
A degenerate-multiplet of Bloch eigenstates at ${\mathbf{Q}}$ can either be considered as a \textit{projective} Rep of $G_{\mathbf{Q}}$, or a linear Rep of $M(\mathbf{Q})$.
Generically, $\tilde{g}$ contains a fractional lattice translation, and the fractional translations and the particle statistics in $\tilde{g}$ uniquely determine the invariants of $G_{\mathbf{Q}}$\cite{Chen1985}.
In sec.\ref{inva} of appendix, we calculate the invariants for each high-symmetry momentum in every type-IV magnetic space group.
In Table \ref{SSGIV}, we list all magnetic space groups with respective high-symmetry momenta where magnons and electrons carry $(\chi_{S},\chi_T,\chi_{PT})=(-1,+1,-1), (+1,-1,+1), (+1,+1,-1), (-1,-1,+1)$, values restricted to magnetic systems.

A final technical remark on terminologies. We use ``quantum numbers'' and ``invariants'' inter-changeably for readability, as the former concept is familiar to non-experts. Rigorously speaking they are certainly not the same: different quantum numbers correspond to different irreducible Reps, while different invariants refer to different classes of Reps. The invariants of $G_\mathbf{Q}$ solely depends on $M$ and $\mathbf{Q}$, and all the Reps at $\mathbf{Q}$ share the same set of invariants.

\appendix\section{Classification of Projective representations of anti-unitary groups $H\times Z_2^T$}%, for $H$ being a point group}
\label{regularProj}

\subsection{Factor Systems and the Invariants}

{\bf Projective representations(Reps) and the factor systems}. As discussed in the main text, the little co-groups of type-IV magnetic space groups have the structure $G=H\times Z_{2}^{T}$ with $Z_{2}^{T}=\{E,\mathbb{T}\},\ \mathbb T^2=E$, where the physical meaning of $\mathbb T$ will be specified later. Supposing $g$ is a group element of $G$, then it is represented by $M(g)$ if $g$ is a unitary element and represented by $M(g)K$ if $g$  is anti-unitary, where $K$ is the complex-conjugate operator satisfying $Ku=u^*K$ with $u$ an arbitrary matrix and $u^*$ its complex conjugation.

The multiplication of (projective) Reps of $g_1, g_2$ depends on if they are unitary or anti-unitary. If we define
\[
s(g)=\left\{
\begin{aligned}
&1,& &{\ \rm if} \ g \ {\rm is\ unitary,\ \ \ }   \\
&-1,& &{\ \rm if}\ g \ {\rm is\ antiunitary,\ \ \ }
\end{aligned}
\right.
\]
and define the corresponding operator $K_{s(g)}$ as
\[
K_{s(g)}=\left\{
\begin{aligned}
&I,& &{\ \rm if\ } s(g)=1,\  \  \\
&K,& &{\ \rm if\ }s(g)=-1,
\end{aligned}
\right.
\]
then we have the multiplication rule of a projective Rep,
%then the above four cases (A)$\sim$(D) can be unified as a single equation
\[
M(g_1)K_{s(g_1)}M(g_2)K_{s(g_2)} = M(g_1g_2)e^{i\theta_2(g_1,g_2)}K_{s(g_1g_2)},
\]
where the $U(1)$ phase factor $\omega_2(g_1,g_2) \equiv e^{i\theta_2(g_1,g_2)}$ is a function of two group variables and is called the {\it factor system}. If $\omega_2(g_1,g_2) = 1$ for any $g_1,g_2\in G$, then above projective Rep is trivial, namely, it is a linear Rep.

Substituting above results into the associativity relation of the sequence of operations $g_1\times g_2\times g_3$, we can obtain
\Beq
&&M(g_1)K_{s(g_1)}M(g_2)K_{s(g_2)}M(g_3)K_{s(g_3)} \ \\
&=& M(g_1g_2g_3)\omega_2(g_1,g_2)\omega_2(g_1g_2,g_3) K_{s(g_1g_2g_3)}\\
&=& M(g_1g_2g_3)\omega_2(g_1,g_2g_3)\omega_2^{s(g_1)}(g_2,g_3)K_{s(g_1g_2g_3)},
\Eeq
namely,
\beq\label{2cocyl}
\omega_2(g_1,g_2)\omega_2(g_1g_2,g_3) = \omega_2^{s(g_1)}(g_2,g_3)\omega_2(g_1,g_2g_3).
\eeq
Eq.(\ref{2cocyl}) is the general relation that the factor systems of any finite group (no matter unitary or anti-unitary) should satisfy.
If we introduce a gauge transformation $M'(g)K_{s(g)}=M(g)\Omega_1(g)K_{s(g)}$, where the phase factor $\Omega_1(g)=e^{i\theta_1(g)}$ depends on a single group variable, then the factor system changes into
\beq\label{gaugeomega2}
\omega'_2(g_1,g_2)=\omega_2(g_1,g_2)\Omega_2(g_1,g_2),
\eeq
with
\beq\label{2cob2}
\Omega_2(g_1,g_2) = {\Omega_1(g_1)\Omega_1^{s(g_1)}(g_2)\over \Omega_1(g_1g_2)}.
\eeq
The equivalent relations (\ref{gaugeomega2}) and (\ref{2cob2}) define the equivalent classes of the solutions of (\ref{2cocyl}). The number of equivalent classes for a finite group is usually finite.

{\bf The  2-cocycles and the 2$^{\rm nd}$ group cohomology.}  The factor systems (\ref{2cocyl}) of projective Reps of group $G$ are also called cocycles. The equivalent classes of the cocycles %associated with the projective Reps of group $G$
form a group, $i.e.$ the group-cohomology group. The group cohomology \footnote{For an introduction to group cohomology, see wiki {\href{http://en.wikipedia.org/wiki/Group_cohomology}{http://en.wikipedia.org/wik/Group\_cohomology}} and
Romyar Sharifi,  {\href{http://citeseerx.ist.psu.edu/viewdoc/summary?doi=10.1.1.296.244}{AN INTRODUCTION TO GROUP COHOMOLOGY}}}%\cite{RS}
$\{{\rm Kernel}\ d/{\rm Image}\ d\}$ is defined by the coboundary operator $d$
%(for details see appendix \ref{app1}),
\begin{eqnarray}
& & (d\omega_{n})(g_{1},\ldots,g_{n+1}) \nonumber\\
&& =[g_{1}\cdot\omega_{n}(g_{2},\ldots,g_{n+1})]\omega^{(-1)^{n+1}}_{n}(g_{1},\ldots,g_{n})\times  \nonumber\\
&&
\prod^{n}_{i=1}\omega^{(-1)^{i}}_{n}(g_{1},\ldots,g_{i-1},g_{i}g_{i+1},g_{i+2},\ldots,g_{n+1}).
\end{eqnarray}
where $g_1,...,g_{n+1}\in G$ and the variables $\omega_{n}(g_{1},\ldots,g_{n})$ take value in an Abelian coefficient group $\mathcal A$ [usually $\mathcal A$ is a subgroup of $U(1)$, in the present work $\mathcal A=U(1)$]. %{\color{blue}
The set of variables $\omega_{n}(g_{1},\ldots,g_{n})$ is called a $n$-cochain. For anti-unitary groups the module $g\cdot$ is defined by
\begin{eqnarray*}\label{module}
g\cdot\omega_{n}(g_{1},\ldots,g_{n}) =\omega_{n}^{s(g)}(g_{1},\ldots,g_{n}).
\end{eqnarray*}

With this notation, Eq. (\ref{2cocyl}) can be rewritten as
\Beq
(d\omega_2)(g_1,g_2,g_3)=1,
\Eeq
the solutions of above equations are called 2-cocycles with $U(1)$-coefficient. Similarly, Eq. (\ref{2cob2}) can be rewritten as
\Beq
\Omega_2(g_1,g_2)=(d\Omega_1)(g_1,g_2),
\Eeq
where $\Omega_1(g_1),\Omega_1(g_2)\in U(1)$ and $\Omega_2(g_1,g_2)$ are called 2-coboundaries. Two 2-cocycles $\omega_2'(g_1,g_2)$ and $\omega_2(g_1,g_2)$ are equivalent if they differ by a 2-coboundary, see Eq. (\ref{gaugeomega2}). The equivalent classes of the 2-cocycles $\omega_2(g_1,g_2)$ form the second group cohomology $\mathcal H^2(G, U(1))$.

Writing  $\omega_{2}(g_{1},g_{2})=e^{i\theta_{2}(g_{1},g_{2})}$, where $\theta_{2}(g_{1},g_{2})\in[0,2\pi)$, then the cocycle equations $(d\omega_{2})(g_{1},g_2,g_{3})=1$ can be written in terms of linear equations,
\begin{eqnarray}
&&s(g_{1})\theta_{2}(g_{2},g_{3})-\theta_{2}(g_{1}g_{2},g_{3})+\theta_{2}(g_{1},g_{2}g_{3})\nonumber\\
&&  -\theta_{2}(g_{1},g_{2})=0.
\label{2cocycle}
\end{eqnarray}
Similarly, if we write $\Omega_1(g_1)=e^{i\theta_1(g_1)}$ and $\Omega_2(g_1,g_2)=e^{i\Theta_2(g_1,g_2)}$, then the 2-coboundary (\ref{2cob2})
can be written as
\beq
\Theta_{2}(g_{1},g_{2})= s(g_{1})\theta_{1}(g_{2})-\theta_{1}(g_{1}g_{2})+\theta_{1}(g_{1}) .
  \label{2coboundary}
\end{eqnarray}
The equal sign in (\ref{2cocycle}) and (\ref{2coboundary}) means equal mod $2\pi$. From these linear equations, we can obtain the solution space of the cocycle equations, as well as the classes that the solutions belong to. The set of classes forms a finite Abelian group, which labels the classification of the projective Reps.

{\bf Invariants of projective Reps.}  The second group-cohomology group is generated by a certain number of {\it invariants}. The invariants are, by definition, invariant under the gauge transformation (\ref{gaugeomega2}) and (\ref{2cob2}). They are formed by independent functions of the cocycles $\omega_{2}(g_{1},g_{2})$.
%The independent gauge invariant variables form the {\it invariants} of the 2-cocycles (or the corresponding projective Reps).
 For instance, for the unitary group $D_2=Z_2\times Z_2=\{E,P\}\times \{E,Q\}$ with $P^2=E, Q^2=E$, the classification of 2-cocycles is
 $$\mathcal H^2(D_2, U(1))=\mathbb Z_2,$$ there  is only one independent invariant which is given by $\chi = {\omega_2(P,Q)\over \omega_2(Q,P)}$. Another example is the simplest anti-unitary group $Z_2^T= \{E,\mathbb{T}\}$, which has $$\mathcal H^2(Z_2^T, U(1))=\mathbb Z_2,$$ %classification of projective Reps,
with the invariant $\chi_T=\omega_2({\mathbb{T}},{\mathbb{T}})$.

 Generally, for any group $G$, if $\mathcal{H}^{2}(G,U(1))=\mathbb Z_{2}^{n}$, then the 2-cocycles of $G$ have $n$ invariants, each taking value $+1$ or $-1$. The invariants of some anti-unitary groups will be given later.

{\bf Irreducible projective Reps.} For the cocycle solution of the each class, we can construct the regular projective Rep, from which all of the inequivalent irreducible projective Reps can be obtained. The method is provided in Ref. \cite{jyzxliu2018} and we will not go to details here.

\subsection{Relation Between the Classification of projective Reps of $H\times Z_2^T$ and that of  of $H$}

The group cohomology of $H\times Z_2^T$ is closely related to that of $H$. Before identifying this relation, we first introduce $Z_2$-coefficient group cohomology and clarify its relation to the $U(1)$-coefficient group cohomology.

{\bf $U(1)$-coefficient and $Z_2$-coefficient}
It can be shown that the $U(1)$-coefficient and $Z_2$-coefficient second cohomology has the following relation,
\beq\label{U1Z2}
\mathcal H^2(G,Z_2) &=& [\mathcal H^2(G,U(1))]_{\mathbb Z_2} \times [\mathcal H^2(G,Z)]_{\mathbb Z_2} \nonumber\\
%,\nonumber\\%\mathcal H^1(G,Z_2),&=&
 &=& [\mathcal H^2(G,U(1))]_{\mathbb Z_2} \times\mathcal H^1(G,Z_2),
\eeq
where $[\mathcal H^2(G,U(1))]_{\mathbb Z_2}=\mathcal H^1(\mathcal H^2(G,U(1)), Z_2)$. The first part on the right-hand side indicates that the $U(1)$-coefficient cohomology group $\mathcal H^2(G,U(1))$ contributes to the $Z_2$-coefficient classification only if it has a $Z_2$ representations; the second part $ [\mathcal H^2(G,Z)]_{\mathbb Z_2}=\mathcal H^1(G,Z_2)$ originates from the collapsing of certain $U(1)$-coboundaries of $G$ when their values are constrained to $Z_2=\{1,-1\}$.

For example, $\mathcal H^2(Z_4\times Z_4,U(1)) = \mathbb Z_4$ and $\mathcal H^1(Z_4\times Z_4,Z_2) = \mathbb Z_2\times \mathbb  Z_2$, therefore
\Beq
\mathcal H^2(Z_4\times Z_4,Z_2) = \mathbb Z_2\times \mathbb Z_2\times \mathbb Z_2,
\Eeq
where one of the $\mathbb Z_2$ comes from $[\mathcal H^2(Z_4\times Z_4,U(1)) ]_{\mathbb Z_2}$ and the  other two $\mathbb Z_2$ come from $\mathcal H^1(Z_4\times Z_4,Z_2)$.

{\bf Classification of $H\times Z_2^T$}. Firstly, for 2-cocycles of any group $H$, we adopt the gauge such that $\omega_2(E, h)=\omega_2(h,E)=1$ for any $h\in H$. Furthermore, for anti-unitary groups $H\times Z_2^T$, %and $H\rtimes Z_2^T$,
one can further fix the gauge such that the following relations are satisfied (see Ref.\cite{jyzxliu2018} for details)
\Beq
&&\omega_2(  {\mathbb{T}},h)=\omega_2(h,  {\mathbb{T}})=1, \\
&&\omega_2(  {\mathbb{T}},  {\mathbb{T}}h)=\omega_2(  {\mathbb{T}}h,  {\mathbb{T}})=\omega_2(  {\mathbb{T}},  {\mathbb{T}}),\\
&&\omega_2(  {\mathbb{T}}h_1,  {\mathbb{T}}h_2)=\omega_2(\bar h_1,h_2)\omega_2(  {\mathbb{T}},  {\mathbb{T}}),\\
&&\omega_2(  {\mathbb{T}}h_1,h_2)=\omega_2^{-1}(h_1,h_2),\\
&&\omega_2(h_1,  {\mathbb{T}}h_2)=\omega_2(h_1,\bar h_2),
\Eeq
with the constraint
\Beq
\omega_2(h_1,h_2)\omega_2(\bar h_1,\bar h_2)=1,
\Eeq
where $\bar h =   {\mathbb{T}}h  {\mathbb{T}}$, $h,h_1,h_2\in H$. Therefore, the classification of the second group cohomology is dependent on $\omega_2(  {\mathbb{T}},  {\mathbb{T}})$ and $\omega_2(h_1,h_2)$. For the direct product group $H\times Z_2^T$, we have $\bar h=h, \omega_2(h_1,h_2)=\pm1$ and consequently
\beq\label{GZ2T}
\mathcal H^2(H\times Z_2^T, U(1))&=&\mathcal H^2(Z_2^T, U(1))\times\mathcal H^2(H, Z_2) \\
%&=& \mathbb Z_2\times \mathcal H^2(H, Z_2)\nonumber\\
&=& \mathbb Z_2\times [\mathcal H^2(H, U(1))]_{\mathbb Z_2}\times \mathcal H^1(H,Z_2),\nonumber
\eeq
where $\mathcal H^1(H,Z_2)$ stands for the twisting between $H$ and $  {\mathbb{T}}$, which corresponds to the general Kramers classes $(g_0  {\mathbb{T}})^2=-1$ for $g_0^2=E$. In the second line of the equation we have used the result (\ref{U1Z2}).

For instance, for the group $G=  {Z_2}\times Z_2^T$, since $\mathcal H^2({Z_2},U(1))=\mathbb Z_1$,$\mathcal H^1({Z_2},{Z_2} )=\mathbb Z_2$, %\del{where $Z_2^P=\{E,P\}$ is the spatial inversion group},
we have
\Beq
&&\mathcal H^2(Z_2\times Z_2^T, U(1))\\
 &&= \mathcal H^2(Z_2^T, U(1)) \times [\mathcal H^2(  {Z_2}, U(1))]_{\mathbb Z_2} \times \mathcal H^1(  {Z_2},Z_2) =\mathbb Z_2^2.
\Eeq

\subsection{Results for abstract and concrete groups}
In Table \ref{H2cohomology}, we give the results of the second group cohomology of several direct product groups $G=H\times Z_{2}^{T}$. All the groups appeared in Table \ref{H2cohomology} are abstract groups. The physical meaning of these groups are specified by mapping these abstract groups into concrete magnetic point groups.
%, which has been illustrated in Tab.\ref{concg}.

In Table \ref{concg}, we list the correspondence between the abstract group $H\times Z_{2}^T$ and the isomorphic magnetic point groups. As before, the abstract group $Z_{2}^{T}$ stands for $Z_{2}^{T}=\{E,\mathbb T\}$, where the anti-unitary generator $\mathbb T$ can be mapped to either $T$ or a product of $T$ with other unitary element. Here we have exclusively denoted $T$ as pure time reversal operation.
The following notations have been used: $C_{2m},C_{4m}^{\pm}(m=x,y,z)$,$C_{3j}^{\pm}(j=1,2,3,4)$,$C_{2},C'_{2i},C''_{2i}(i=1,2,3)$,$C_{2p}(p=a,b,c,d,e,f)$,see Fig 1.1,1.2 and 1.3 of Ref.\cite{Bradley2010}, $P={\rm space\ inversion}$, $\mathcal{M}=$ mirror reflection plane, $PC_{2}=\mathcal{M}_{h}$,$PC_{2m}=\mathcal{M}_{m}$,$PC'_{2i}=\mathcal{M}_{di}$,$PC''_{2i}=\mathcal{M}_{vi}$, $PC_{2p}=\mathcal{M}_{dp}$,$PC_{3}^{\pm}=S_{6}^{\mp}$, $PC_{4m}^{\pm}=S_{4m}^{\mp}$,$PC_{6}^{\pm}=S_{3}^{\mp}$,$PC_{3j}^{\pm}=S_{6j}^{\mp}$.

The definitions of invariants of the second group cohomology for the concrete groups are also given in Table \ref{concg}.  Notice that for $G_{\mathbf{Q}}=D_{2h}\times Z_{2}^T,D_{4h}\times Z_{2}^T,D_{6h}\times Z_{2}^T,T_{h}\times Z_{2}^T,O_{h}\times Z_{2}^T$, the first three invariants correspond to the quantum numbers $(\chi_S,\chi_T,\chi_{PT})$ that were discussed in main text.

\section{Invariants for the little co-groups in type-IV magnetic space groups}\label{inva}

For a Shubnikov magnetic space group $M$, %also called magnetic space group,
the magnetic little group $M(\mathbf{Q})$ is the subgroup of $M$ which transforms the wave vector $\mathbf{Q}$ in the first Brillouin zone (BZ)  into its equivalent wave vector $\mathbf{Q}+\mathbf {K}$, here $\mathbf K$ is reciprocal lattice vector. $M(\mathbf{Q})$ is also a Shubnikov space group which has the translation group as its normal subgroup. The point group of $M(\mathbf{Q})$ is called magnetic little co-group $G_{\mathbf{Q}}$, which is the quotient group of $M(\mathbf{Q})$ with respect to the translation group as discussed in the main text.

A type \textrm{IV} Shubnikov space group \cite{Bradley2010} has the form $M=H+T\{E|\mathbf t_0\}H $, where $H$ belongs to the 230 (Fedorov) space groups and $\mathbf t_0$ is the extra fractional translation about the time-reversal operation $T$. The magnetic little co-group of $M$  at point $\mathbf Q$ is isomorphic to the direct product of a point group $H$ and $Z_{2}^T$, namely,  $G_{\mathbf{Q}}=H\times Z_{2}^T$, where $Z_{2}^T=\{E,T\}$ [if $T$ is an element of the magnetic point group of $M(\mathbf Q)$, or equivalently $\tilde T\in M(\mathbf{Q})$, where $\tilde T$ is defined in (\ref{gtilde}) below] or $Z_{2}^T=\{E,PT\}$ (if $\tilde T\notin M(\mathbf Q)$) depending on the symmetry of the high symmetry point (also see Table \ref{concg}). For all the type \textrm{IV} magnetic space groups of our interest whose point group contains $P$, we always have $PT\in G_{\mathbf Q}$.

As illustrated in the main text, the linear Rep of the Shubnikov space group $M$ at some high symmetry point  $\mathbf Q$ (or on a high symmetry line) defines a projective Rep of the little co-group $G_{\mathbf Q}$. Now we derive the details here. For any group element $g\in G_{\mathbf Q}$, we denote
\beq\label{gtilde}
\tilde g= \{g | \mathbf t_g\} \in M,
\eeq
where $\mathbf t_g = x_g\mathbf a + y_g\mathbf b + z_g\mathbf c$ is the unique fractional translation associated with $g$, with $\mathbf{a,b,c}$ the basis vectors of the lattice and $0\leq x_g,y_g,z_g<1$. Then we further denote $\hat g$ as the linear Rep matrix of $\tilde g\in M$. Then for $\tilde g_1, \tilde g_2 \in M$, we have the group multiplication rule (for simplicity, we temporarily assume that both of $\tilde g_1, \tilde g_2$ are unitary operations)
\Beq
\tilde g_1\tilde g_2 &=& \{g_1|\mathbf t_{g_1}\}\{g_2|\mathbf t_{g_2}\} =\{g_1g_2 | g_1\mathbf t_{g_2} +\mathbf t_{g_1}\}\\
&=& \{g_1g_2 |\mathbf t_{g_1g_2}\} \omega(g_1,g_2),
\Eeq
where $ \omega(g_1,g_2)=\{E| g_1\mathbf t_{g_2} +\mathbf t_{g_1} - \mathbf t_{g_1g_2}\}$ is an element of the translation group. Accordingly, we have
\[
\hat g_1\hat g_2 = \widehat {g_1g_2}\hat \omega(g_1,g_2) %= \widehat {g_1g_2}e^{-i\mathbf Q\cdot (g_1\mathbf t_{g_2} +\mathbf t_{g_1} - \mathbf t_{g_1g_2})}
\]
where $\hat\omega(g_1,g_2) = e^{-i\mathbf Q\cdot (g_1\mathbf t_{g_2} +\mathbf t_{g_1} - \mathbf t_{g_1g_2})}$. Therefore, if we treat $\hat g$ as a projective Rep of $g\in G_{\mathbf Q}$, then $\hat\omega(g_1,g_2)$ defines the factor system. It can be verified that thus defined factor system satisfy the 2-cocycle equation. On the other hand, if we know the projective Rep of $G_{\mathbf Q}$ with the factor system $\hat\omega(g_1,g_2)$ defined above, then we can easily obtain the linear Rep of the total magnetic space group $M$.

%Owing to the fractional translations in the non-symmorphic groups, a factor system is obtained for the little co-group $G_{\mathbf{Q}}$ \cite{BradleyCrack,ChenJQRMP85,ChenJQ02},
Actually, up to a gauge transformation, the factor system mentioned above can be obtained in a systematic way\cite{Bradley2010,Chen1985,chen2002group}. In the following we generalize the rule introduced in Ref.\cite{Bradley2010,Chen1985,chen2002group} to the anti-unitary groups ({\it i.e.} magnetic space groups). For arbitrary elements  $g_1,g_2\in G_{\bf Q}$ (and accordingly $\tilde g_1, \tilde g_2\in M$), we have the factor system
\begin{equation}\label{phasemag}
\omega_2(g_1,g_2)=e^{-i\mathbf K_{g_1}\cdot\mathbf t_{g_2}},
\end{equation}
where %$g_1,g_2\in  G_{\mathbf{Q}}$,
$\mathbf K_{g_1}=s(g_1)(g_1^{-1}\mathbf Q -\mathbf Q)$ is a reciprocal lattice vector and $\mathbf t_{g_2}$ is the fractional translation associated with $g_{2}$. %\in G_{\mathbf{Q}}$.
For any  $g_{1},g_{2},g_{3}\in G_{\mathbf{Q}}$, one can verify that
the cocycle equation (\ref{2cocyl}) is satisfied,
\begin{eqnarray}
&& \omega_2(g_1,g_2 g_3)\omega_2^{s(g_1)}(g_2,g_3) \nonumber\\
&=&e^{-i\mathbf K_{g_1}\cdot\mathbf t_{g_2 g_3}}e^{-is(g_{1})\mathbf K_{g_2}\cdot\mathbf t_{g_3}}    \nonumber\\
&=&e^{-i\mathbf K_{g_1}\cdot\mathbf t_{g_2}}e^{-i\mathbf K_{g_1}\cdot g_{2}\mathbf t_{g_3}}
e^{-is(g_{1})\mathbf K_{g_2}\cdot\mathbf t_{g_3}}    \nonumber\\
&=&e^{-i\mathbf K_{g_1}\cdot\mathbf t_{g_2}}e^{-is(g_2)g_{2}^{-1}\mathbf K_{g_1}\cdot \mathbf t_{g_3}}
e^{-is(g_{1})\mathbf K_{g_2}\cdot\mathbf t_{g_3}}    \nonumber\\
&=&e^{-i\mathbf K_{g_1}\cdot\mathbf t_{g_2}}e^{-is(g_2)s(g_1)[(g_{1}g_{2})^{-1}\mathbf Q-g_{2}^{-1}\mathbf Q]
\cdot \mathbf t_{g_3}} \nonumber\\
&&
\cdot e^{-is(g_{1})s(g_2)(g_2^{-1}\mathbf Q -\mathbf Q)\cdot\mathbf t_{g_3}}    \nonumber\\
&=&e^{-i\mathbf K_{g_1}\cdot\mathbf t_{g_2}}
e^{-is(g_{1}g_{2})[(g_{1}g_{2})^{-1}\mathbf Q-\mathbf Q]\cdot\mathbf t_{g_3}}\nonumber\\
&=&e^{-i\mathbf K_{g_1}\cdot\mathbf t_{g_2}}e^{-i\mathbf K_{g_{1}g_{2}}\cdot\mathbf t_{g_3}} \nonumber\\
&=&\omega_2(g_1,g_2)\omega_2(g_{1}g_{2},g_{3}),
\end{eqnarray}
here $\mathbf t_{g_{2}g_{3}}=\mathbf t_{g_2}+g_2 \mathbf t_{g_3}+\mathbf R_{23}$, $\mathbf K_{g_1}\cdot g_{2}\mathbf t_{g_3}=s(g_2)g_{2}^{-1}\mathbf K_{g_1}\cdot \mathbf t_{g_3}$ and $\mathbf R_{23}$ is a lattice vector.

For bosonic particles whose bare spin $s_0$ is an integer, the band structure is characterized by projective Rep of $G_{\mathbf{Q}}$ with the factor system (\ref{phasemag}), now we note it as $\omega_{2b}(g_1,g_2)$. The bosonic particles are characterized by the invariant $\chi_{PT}=\omega_2(PT,PT)=(-1)^{2s_0}=1$ (this is no surprise because $(PT)^{-1}\mathbf Q=\mathbf Q$ such that $\mathbf K_{PT}$=0). The magnon excitations in magnetic ordered ground states belong to this case.

For fermionic particles whose bare spin $s_0$ is a half-odd-integer,  an extra factor system $\omega_2^{({1\over2})}(g_1,g_2)$ is contributed from the spin rotation (owing to spin-orbit coupling), which corresponds to the double valued Reps of the little co-group. The non-trivial invariants of factor system $\omega_2^{({1\over2})}(g_1,g_2)$ are $\chi_{S}=\frac{\omega_{2}(C_{2m},C_{2n})}{\omega_{2}(C_{2n},C_{2m})}=-1$,$\chi_{T}=\omega_2(T,T)=-1$,
$\chi_{PT}=\omega_2(PT,PT)=-1$,where $m\neq{n}$ take values $x,y,z$. Combining with the factor system from fractional translation, the total factor system for fermions is given by $$\omega_{2f}(g_1,g_2)=\omega_2^{({1\over2})}(g_1,g_2) \omega_{2b}(g_1,g_2).$$ The fermionic particles (such as electrons) are characterized by the invariant $\chi_{PT}=-1$.

From the given factor systems and the definition of the invariants {in Table \ref{concg}}, we can easily obtain the values of the invariants. In Table \ref{SSGIV}, we list the little co-groups of the high symmetry points of  type \textrm{IV} Shubnikov space groups, and also provide the values of the invariants for the two different classes of factors system discussed above.  The values $(-1,+1,-1),(+1,-1,+1), (+1,+1,-1), (-1,-1,+1)$ of $(\chi_S,\chi_T,\chi_{PT})$ are only accessible in magnetic materials, which are marked with red for electrons, and blue for magnons.

\section{$SO(4)$ algebra and cross-resonant absorption under symmetry reduction}\label{sysredu}

In this section we derive a few statements made in the main article, with respect to whether the $SO(4)$ algebra, among polarization and magnetization operators for the $R_{1/2}$-Rep, as well as the cross-resonant absorption phenomenon, in 222.103 is preserved or broken, if the little co-group is broken from $O_h\times{Z}_{2}^{T}$ to $T_h\times{Z}_{2}^{T}$, $D_{4h}\times{Z}_{2}^{T}$ and $D_{2h}\times{Z}_{2}^{T}$.
These symmetry reduction naturally comes into consideration as one (i) breaks the fourfold axis into a twofold axis ($T_h$), (ii) breaks the threefold axis ($D_{4h}$), and (iii) both the fourfold and the threefold axis ($D_{2h}$), leaving the parity, time-reversal, and three twofold axes intact.

Our first statement is that, after the reduction, $R_{1/2}$ remains an irreducible Rep for the little co-group.
In all three scenarios, $D_{2h}\times{Z}_{2}^{T}$ is a subgroup of the little co-group, therefore, the three invariants $\chi_{S,T,PT}$ simply ``inherit'' their respective values in 222.103, $(\chi_{S},\chi_{T},\chi_{PT})=(-1,+1,-1)$, because these three invariants only involve parity, time-reversal, and twofold axes, all inside $D_{2h}\times{Z}_{2}^{T}$.
From Eq.\eqref{eq:5} in main text, we have
\begin{eqnarray}
\label{eq:S1}\hat{T}&=&\tau_y\sigma_y{K},\\
\label{eq:S2}\hat{P}\hat{T}&=&i\sigma_y{K},\\
\label{eq:S3}\hat{C}_{2x,2y,2z}&=&i\sigma_{x,y,z}.
\end{eqnarray}
From Eq.(\ref{eq:S3}), we see that mass terms (matrices that commute with all symmetris) must take the form $\tau_{\mu=0,1,2,3}\sigma_0$, yet Eq.(\ref{eq:S1}) narrows these possibilities down to $\tau_0\sigma_0$.
A constant mass term cannot open a gap, so $R_{1/2}$ remains an irreducible Rep under all three types of symmetry reduction.

The next statement is that for $O_h\rightarrow{T}_h$, the $SO(4)$ algebra among polarization and magnetization operators still holds.
We make use of Table \ref{Tab:1} to simplify our derivation.
We observe that as $O_h\rightarrow{T}_h$, the Reps such as $A_{1u}^+$ and $A_{2u}^+$ become identical, because their difference solely appears in how they transform under $C_4$, broken in $T_h$.
Similarly, $T_{1u}^+$ and $T_{2u}^+$ become the same $T_u^+$.
The electric-field components $(E_x,E_y,E_z)$ form a $T_u^+$-Rep, so there are two linearly independent sets of $\Gamma$-matrices that couple to the electric field:
\begin{eqnarray}
\hat{d}^{(1)}_{i=x,y,z}&=&\tau_z\sigma_i/2,\\
\nonumber
\hat{d}^{(2)}_{i=x,y,z}&=&\tau_x\sigma_i/2.
\end{eqnarray}
The physical (reduced) polarization $\hat{p}$ is a linear superposition of $\hat{d}^{(1)}$ and $\hat{d}^{(2)}$:
\begin{equation}\label{eq:S5}
\hat{p}_i=\hat{d}^{(1)}_i\cos\theta+\hat{d}^{(2)}_i\sin\theta.
\end{equation}

On the other hand, while $T_{1g,2g}^-\rightarrow{T}_g^-$ as $O_h\rightarrow{T}_h$, according to Table \ref{Tab:1} there is not a $T_{2g}^-$ for all 16 $\Gamma$-matrices, so $T_g^-$-matrices are just the $T_{1g}^-$-matrices in Table \ref{Tab:1}.
Physically, this means that there is not any other choice for the (reduced) magnetization:
\begin{equation}\label{eq:S6}
\hat{m}_i=\tau_0\sigma_i/2.
\end{equation}
It is obvious from Eq.(\ref{eq:S5},\ref{eq:S6}) that $\hat{p}_i,\hat{m}_i$ still satisfy Eq.\eqref{eq:9} in the main text, i.e., the $SO(4)$ algebra is preserved.

Our third statement is that as symmetry is reduced to $D_{4h}$, the $SO(4)$ algebra is broken.
We show this, again, by using Table \ref{Tab:1} and the reduction of irreducible Rep from $O_h$ to $D_{4h}$: $T_{1u}^+\rightarrow{A_{2u}}\oplus{E_u}, T_{2u}^+\rightarrow{B_{2u}}\oplus{E_u}$.
The electric-field component $E_z$ belongs to $A_{2u}$ and $(E_x,E_y)$ belong to $E_u$.
Therefore there is only one choice for $\hat{p}_z$, yet two choices for $\hat{p}_{x,y}$:
\begin{eqnarray}\label{eq:S7}
\hat{p}_z&=&\tau_z\sigma_z/2,\\
\nonumber
\hat{p}_{x,y}&=&\hat{d}^{(1)}_{x,y}\cos\theta+\hat{d}^{(2)}_{x,y}\sin\theta.
\end{eqnarray}
Since there is not a $T_{2g}^-$-Rep in Table \ref{Tab:1}, the choice for $\hat{m}_i$ remains the same.
Obviously, the $\hat{p}_i$ operators in Eqs.(\ref{eq:S7}) do not form an $SU(2)$ algebra, so the $SO(4)$ algebra is also broken.
For the $D_{2h}$ scenario, the derivation is similar to the $T_h$ case, with the only difference that, in $D_{2h}$, the $\theta$ in Eq.(\ref{eq:S5}) becomes three different values $\theta_{x,y,z}$ for three components.
\begin{equation}\label{eq:S8}
\hat{p}_i=\hat{d}^{(1)}_i\cos\theta_i+\hat{d}^{(2)}_i\sin\theta_i.
\end{equation}
Due to the component-dependent $\theta_i$, $\hat{p}_i$ do not form an $SU(2)$ algebra, and the $SO(4)$ algebra is again broken.

Our final statement is that even in the absence of $SO(4)$ algebra, the cross-resonant absorption still holds under symmetry reduction.
For this purpose, we only need to show the absorption in $D_{2h}$-scenario, because it has the lowest symmetry.
Consider an $\hat{m}_z$-polarized state $\psi_0$: $\hat{m}_{i}\psi_0=\delta_{iz}\psi_0$.
Using the explicit expression of $\hat{m}_z$, we have
\begin{equation}
\psi=u|11\rangle+v|\bar11\rangle,
\end{equation}
where $|mn\rangle$ denotes an eigenstate of $\tau_z,\sigma_z$ of eigenvalues $m$ and $n$, respectively.
Now consider an electric field along $x$-axis, so that the Hamiltonian is
\begin{equation}
\hat{h}=\lambda_{E,x}(\tau_z\sigma_x\cos\theta_x+\tau_x\sigma_x\sin\theta_x).
\end{equation}
At $t>0$, we have
\begin{widetext}
\begin{eqnarray}\label{eq:S11}
\psi(t)&=&e^{-i\hat{h}t}\psi_0\\
\nonumber
&=&\cos(\lambda_{E,x}t)\psi_0-i\sin(\lambda_{E,x}t)(\tau_z\sigma_x\cos\theta_x+\tau_x\sigma_x\sin\theta_x)\psi_0\\
\nonumber
&=&\cos(\lambda_{E,x}t)(u|11\rangle+v|\bar11\rangle)-i\sin(\lambda_{E,x}t)\cos\theta_x(u|1\bar1\rangle-v|\bar1\bar1\rangle)
-i\sin(\lambda_{E,x}t)\sin\theta_x(u|\bar1\bar1\rangle+v|1\bar1\rangle)\\
\nonumber
&=&\cos(\lambda_{E,x}t)(u|11\rangle+v|\bar11\rangle)-i\sin(\lambda_{E,x}t)(u\cos\theta_x+v\sin\theta_x)|1\bar1\rangle
-i\sin(\lambda_{E,x}t)(u\sin\theta_x-v\cos\theta_x)|\bar1\bar1\rangle.
\end{eqnarray}
\end{widetext}
Eq.(\ref{eq:S11}) shows a clear oscillation between $\psi_0$ and an orthogonal $\psi'_0=(u\cos\theta_x+v\sin\theta_x)|1\bar1\rangle+(u\sin\theta_x-v\cos\theta_x)|\bar1\bar1\rangle$, with period $T=\pi/\lambda_{E,x}$.
Therefore, if an incident light is propagating along $y$-axis and polarized along $x$-axis, it has an absorption peak at $\omega_0=2\pi/T=2\lambda_{E,x}$.

\section{The other two Reps of $O_{h}\times Z_{2}^T$ at $(\pi/a,\pi/a,\pi/a)$ in 222.103}\label{3/2reps}
% \ref{eq:5}
In addition to Eq.\eqref{eq:5}, there exist another two inequivalent $4$-dim irreducible Reps $R^{\pm}_{3/2}$.
By using the three spin-$\frac{3}{2}$ angular momentum operators
\begin{eqnarray}
J_{x}&=&\frac{1}{2}\left(\begin{array}{cccc}
0        & \sqrt{3} & 0        & 0       \\
\sqrt{3} &  0       & 2        & 0       \\
0        &  2       & 0        & \sqrt{3}\\
0        &  0       & \sqrt{3} & 0
\end{array}\right),
\end{eqnarray}
\begin{eqnarray}
J_{y}&=& -\frac{i}{2}\left(\begin{array}{cccc}
0        & \sqrt{3} & 0        & 0       \\
-\sqrt{3}&  0       & 2        & 0       \\
0        & -2       & 0        & \sqrt{3}\\
0        &  0       & -\sqrt{3}& 0
\end{array}\right), \\
J_{z}&=& \frac{1}{2}\left(\begin{array}{cccc}
3 & 0 & 0 & 0 \\
0 & 1 & 0 & 0 \\
0 & 0 &-1 & 0 \\
0 & 0 & 0 &-3
\end{array}\right),
\end{eqnarray}
they are written as
\begin{eqnarray}
{\hat{T}}&=&\tau_{z}\sigma_{x}K, \nonumber\\
  \hat{P} &=&i\tau_{y}\sigma_{z}, \nonumber\\
  \hat{C}_{2x} &=&\exp(iJ_{x}\pi)=-i\tau_{x}\sigma_{x}, \nonumber\\
  \hat{C}_{31}^{}&\equiv &\hat{C}_{3,111}^{}=-\exp(-\frac{2\pi iJ_{111}}{3}),  \nonumber\\
  \hat{C}_{4z}^{}&=&-i\exp(-\frac{iJ_{z}\pi}{2})=-\frac{1}{\sqrt{2}}(\tau_{z}\sigma_{0}-i\tau_{z}\sigma_{z});
  \label{eqrep36}
\end{eqnarray}
and
\begin{eqnarray}
{\hat{T}}&=&\tau_{z}\sigma_{x}K, \nonumber\\
  \hat{P} &=&-i\tau_{y}\sigma_{z}, \nonumber\\
  \hat{C}_{2x} &=&\exp(iJ_{x}\pi)=-i\tau_{x}\sigma_{x}, \nonumber\\
  \hat{C}_{31}^{}&\equiv &\hat{C}_{3,111}^{}=-\exp(-\frac{2\pi iJ_{111}}{3}),  \nonumber\\
  \hat{C}_{4z}^{}&=&i\exp(-\frac{iJ_{z}\pi}{2})=\frac{1}{\sqrt{2}}(\tau_{z}\sigma_{0}-i\tau_{z}\sigma_{z}),
\label{eqrep64}
\end{eqnarray}
here $J_{111}\equiv\frac{J_{x}+J_{y}+J_{z}}{\sqrt{3}}$. The two Reps (\ref{eqrep36}) and (\ref{eqrep64}) have quadratic ($T_{2g}^{+}$) and quintic ($E_{u}^{-}$) dispersion terms. They can couple electric and magnetic fields, but the minimal coupling terms about electric field do not have the $SO(4)$ algebra as Eq.\eqref{eq:9}.
Even so, all the three inequivalent Reps \eqref{eq:5}, (\ref{eqrep36}) and (\ref{eqrep64}) can lead to cross-resonant absorption. Furthermore, when the above two Reps are reduced from $O_h\times{Z}_{2}^{T}$ to $T_h\times{Z}_{2}^{T}$, $D_{4h}\times{Z}_{2}^{T}$ and $D_{2h}\times{Z}_{2}^{T}$, the cross-resonant absorption also appears, which will be discussed in the future.

\begin{table*}[t]
\caption{ Some direct product groups $G=H\times Z_{2}^T$ and their 2nd group cohomology which classify their projective Reps.
} \label{H2cohomology}
\centering
\begin{tabular}{ |c|c||c|c||c|c| }
\hline
Abstract group $G$& $\mathcal{H}^{2}(G,U(1))$ &Abstract group $G$& $\mathcal{H}^{2}(G,U(1))$
&Abstract group $G$& $\mathcal{H}^{2}(G,U(1))$\\
\hline
$ Z_{2}^T$ & $\mathbb{Z}_{2}$ &

$(Z_{4}\times Z_{2})\times Z_{2}^T$&  $\mathbb{Z}_{2}^{4}$ &

$(Z_{6}\times Z_{2})\times Z_{2}^T$& $\mathbb{Z}_{2}^{4}$   \\
\hline

$Z_{2}\times Z_{2}^T$ & $\mathbb{Z}_{2}^{2}$ &

$[(Z_4\rtimes Z_2)\times Z_{2}]\times Z_2^T$& $\mathbb{Z}_{2}^{7}$ &

$[(Z_6\rtimes Z_2)\times Z_{2}]\times Z_2^T$& $\mathbb{Z}_{2}^{7}$   \\
\hline

$(Z_{2}\times Z_{2})\times Z_{2}^T$ & $\mathbb{Z}_{2}^{4}$ &

$Z_{3}\times Z_{2}^T$&$\mathbb{Z}_{2}$ &

$A_{4}\times Z_{2}^T$&$\mathbb{Z}_{2}^{2}$   \\
\hline

$[(Z_{2}\times Z_{2})\times Z_{2}]\times Z_{2}^T$ & $\mathbb{Z}_{2}^{7}$ &

$Z_{6}\times Z_{2}^T$&$\mathbb{Z}_{2}^{2}$ &

$(A_{4}\times Z_{2})\times Z_{2}^T$& $\mathbb{Z}_{2}^{3}$   \\
\hline

$Z_{4}\times Z_{2}^T$ & $\mathbb{Z}_{2}^{2}$ &

$(Z_{3}\rtimes Z_{2})\times Z_{2}^T$&$\mathbb{Z}_{2}^{2}$ &

$(A_{4}\rtimes Z_{2})\times Z_{2}^T$& $\mathbb{Z}_{2}^{3}$   \\
\hline

$(Z_{4}\rtimes Z_{2})\times Z_{2}^T$ & $\mathbb{Z}_{2}^{4}$ &

$(Z_{6}\rtimes Z_{2})\times Z_{2}^T$& $\mathbb{Z}_{2}^{4}$&

$[(A_{4}\rtimes Z_{2})\times Z_{2}]\times Z_{2}^T$& $\mathbb{Z}_{2}^{5}$   \\
\hline
\end{tabular}
\end{table*}

\begin{table*}[htbp]
\caption{Correspondence between abstract groups and concrete magnetic point groups, where the physical meaning of $\mathbb T$ (the generator of $Z_{2}^T$) is specified as the last generator as listed in the third column. In the fourth column, we list all the invariants of the second group cohomology. Here,$\chi_T\equiv\omega_{2}(T,T)$,$\chi_{PT}\equiv\omega_{2}(PT,PT)$, $\chi_{TC_{2}}\equiv\omega_{2}(TC_{2},TC_{2})$,$\chi_{T\mathcal{M}}\equiv\omega_{2}(T\mathcal{M},T\mathcal{M})$.
} \label{concg}
\centering
\begin{tabular}{ |c|c|c|c| }
\hline
Abstract group& Concrete group& Generators  & Invariants\\ %\del{of point group}
\hline
$Z_{2}\times Z_{2}^T$
&$C_{i}\times Z_{2}^T$
& $P, T$      %$C_{i}=\{E, P\}$
& ($\chi_T$,$\chi_{PT}$)\\
\hline

$(Z_{2}\times Z_{2})\times Z_{2}^T$
&$C_{2v}^{}\times Z_{2}^T$
&   $C_{2x},\mathcal{M}_{y}, PT$     %$C_{2v}^{}=\{E,C_{2x},\mathcal{M}_{y},\mathcal{M}_{z}\}$
&($\frac{\omega_{2}(\mathcal{M}_{y},C_{2x})}{\omega_{2}(C_{2x},\mathcal{M}_{y})}$,$\chi_{PT}$,
$\chi_{T\mathcal{M}_{x}}$, $\chi_{TC_{2y}}$)  \\

%&$C_{2v}^{2}\times Z_{2}^T$, $C_{2v}^{2}=\{E,\mathcal{M}_{x}, C_{2y},\mathcal{M}_{z}\}$,$\mathbb{T}=PT$  \\
%
%&$C_{2v}^{3}\times Z_{2}^T$, $C_{2v}^{3}=\{E,\mathcal{M}_{x}, \mathcal{M}_{y},C_{2z}\}$,$\mathbb{T}=PT$  \\
%
%&$C_{2v}^{4}\times Z_{2}^T$, $C_{2v}^{4}=\{E,\mathcal{M}_{db}, \mathcal{M}_{da},C_{2z}\}$,$\mathbb{T}=PT$  \\
%
%&$C_{2v}^{5}\times Z_{2}^T$, $C_{2v}^{5}=\{E, \mathcal{M}_{d1},\mathcal{M}_{v1}, C_{2}\}$,$\mathbb{T}=PT$  \\
%
%&$C_{2v}^{6}\times Z_{2}^T$, $C_{2v}^{6}=\{E, C'_{21},\mathcal{M}_{v1},\mathcal{M}_{h}\}$,$\mathbb{T}=PT$  \\
%
%&$C_{2v}^{7}\times Z_{2}^T$, $C_{2v}^{7}=\{E,\mathcal{M}_{d1},C''_{21}, \mathcal{M}_{h}\}$,$\mathbb{T}=PT$  \\
%
%&$C_{2v}^{8}\times Z_{2}^T$, $C_{2v}^{8}=\{E,\mathcal{M}_{d2},C''_{22}, \mathcal{M}_{h}\}$,$\mathbb{T}=PT$  \\
%
%&$C_{2v}^{9}\times Z_{2}^T$, $C_{2v}^{9}=\{E,\mathcal{M}_{db},C_{2a}, \mathcal{M}_{z}\}$,$\mathbb{T}=PT$  \\
%
%&$C_{2v}^{10}\times Z_{2}^T$, $C_{2v}^{10}=\{E,C_{2b}, \mathcal{M}_{da},\mathcal{M}_{z}\}$,$\mathbb{T}=PT$  \\
%
%&$C_{2v}^{11}\times Z_{2}^T$, $C_{2v}^{11}=\{E,C_{2c}, \mathcal{M}_{y},\mathcal{M}_{de}\}$,$\mathbb{T}=PT$  \\

&$D_{2}^{}\times Z_{2}^T$
&$C_{2x}, C_{2y}, PT$     %$D_{2}^{}=\{E,C_{2x}, C_{2y},C_{2z}\}$
&($\frac{\omega_{2}(C_{2y},C_{2x})}{\omega_{2}(C_{2x},C_{2y})}$,$\chi_{PT}$,
$\chi_{T\mathcal{M}_{x}}$, $\chi_{T\mathcal{M}_{y}}$)  \\
\hline

\end{tabular}
\end{table*}

\clearpage
\normalem
\bibliography{ref}

\begin{table*}[htbp]
\caption{All invariants of magnetic point groups $H\times Z_{2}^T$ and symmetry points in Brillouin zone to realize them in type IV Shubnikov space groups with $P={\rm space\ inversion}$, $T={\rm time\ reversal}$ symmetries.}\label{SSGIV}
\centering
\begin{tabular}{ |c|cc|c|c| }
\hline
Shubnikov space group &symmetry point& $G_{\mathbf{Q}}$ &invariants (electron)& invariants (Magnon)\\
\hline
2.7~\textrm{IV}&$\Gamma$&$C_{i}\times Z_{2}^T$&($-1,-1$)  &($+1,+1$) \\

&B&$C_{i}\times Z_{2}^T$&($+1,-1$)  &($-1,+1$) \\

&F&$C_{i}\times Z_{2}^T$&($-1,-1$)  &($+1,+1$) \\

&G&$C_{i}\times Z_{2}^T$&($-1,-1$)  &($+1,+1$) \\
\hline
10.47~\textrm{IV}&$\Gamma$&$C_{2h}^{1}\times Z_{2}^T$&($+1,-1,+1,-1$)  &($+1,+1,+1,+1$) \\

&B&$C_{2h}^{1}\times Z_{2}^T$&($+1,-1,+1,-1$)  &($+1,+1,+1,+1$) \\

&Y&$C_{2h}^{1}\times Z_{2}^T$&($+1,+1,+1,-1$)  &($+1,-1,+1,+1$) \\
\hline

\end{tabular}
\end{table*}

\begin{table*}[htbp]
(Continued from the previous page)\\
\centering
\begin{tabular}{ |c|cc|c|c| }
\hline
Shubnikov space group &symmetry point& $G_{\mathbf{Q}}$ &invariants (electron)& invariants (Magnon)\\
\hline

12.64~\textrm{IV}&$\Gamma$&$C_{2h}^{1}\times Z_{2}^T$&($+1,-1,+1,-1$)  &($+1,+1,+1,+1$) \\

&A&$C_{2h}^{1}\times Z_{2}^T$&($+1,-1,+1,-1$)  &($+1,+1,+1,+1$)\\

&Z&$C_{2h}^{1}\times Z_{2}^T$&($+1,-1,+1,-1$)  &($+1,+1,+1,+1$) \\

&M&$C_{2h}^{1}\times Z_{2}^T$&($+1,-1,+1,-1$)  &($+1,+1,+1,+1$)\\

&L&$C_{i}\times Z_{2}^T$&($+1,-1$)  &($-1,+1$) \\

&V&$C_{i}\times Z_{2}^T$&($+1,-1$)  &($-1,+1$) \\
\hline

13.70~\textrm{IV}&$\Gamma$&$C_{2h}^{1}\times Z_{2}^T$&($+1,-1,+1,-1$)  &($+1,+1,+1,+1$) \\
\hline

\end{tabular}
\end{table*}

\begin{table*}[htbp]
(Continued from the previous page)\\
\centering
\begin{tabular}{ |c|cc|c|c| }
\hline
Shubnikov space group &symmetry point& $G_{\mathbf{Q}}$ &invariants (electron)& invariants (Magnon)\\
\hline

14.82~\textrm{IV}&$\Gamma$&$C_{2h}^{1}\times Z_{2}^T$&($+1,-1,+1,-1$)  &($+1,+1,+1,+1$)\\

&B&$C_{2h}^{1}\times Z_{2}^T$&($-1,+1,+1,-1$)  &($-1,-1,+1,+1$) \\

&Y&$C_{2h}^{1}\times Z_{2}^T$&($+1,-1,+1,-1$)  &($+1,+1,+1,+1$)\\

&Z&$C_{2h}^{1}\times Z_{2}^T$&($-1,-1,-1,-1$)  &($-1,+1,-1,+1$) \\

&C&$C_{2h}^{1}\times Z_{2}^T$&($-1,-1,-1,-1$)  &($-1,+1,-1,+1$)  \\

&D&$C_{2h}^{1}\times Z_{2}^T$&($+1,+1,-1,-1$)  &($+1,-1,-1,+1$) \\

&A&$C_{2h}^{1}\times Z_{2}^T$&($-1,+1,+1,-1$)  &($-1,-1,+1,+1$) \\
\hline

\end{tabular}
\end{table*}

\begin{table*}[htbp]
(Continued from the previous page)\\
\centering
\begin{tabular}{ |c|cc|c|c| }
\hline
Shubnikov space group &symmetry point& $G_{\mathbf{Q}}$ &invariants (electron)& invariants (Magnon)\\
\hline

48.262~\textrm{IV}&$\Gamma$&$D_{2h}^{1}\times Z_{2}^T$&($-1,-1,-1,+1,+1,+1,+1$)  &($+1,+1,+1,+1,+1,+1,+1$) \\

&Y&$D_{2h}^{1}\times Z_{2}^T$&($-1,-1,-1,+1,+1,-1,+1$)  &($+1,+1,+1,+1,+1,-1,+1$) \\

&X&$D_{2h}^{1}\times Z_{2}^T$&(${\color{red}-1,+1,-1},-1,+1,+1,-1$)  &(${\color{blue}+1,-1,+1},-1,+1,+1,-1$) \\

&Z&$D_{2h}^{1}\times Z_{2}^T$&($-1,-1,-1,+1,+1,-1,-1$)  &($+1,+1,+1,+1,+1,-1,-1$) \\

&U&$D_{2h}^{1}\times Z_{2}^T$&(${\color{red}-1,+1,-1},-1,+1,-1,+1$)  &(${\color{blue}+1,-1,+1},-1,+1,-1,+1$) \\

&T&$D_{2h}^{1}\times Z_{2}^T$&($-1,-1,-1,+1,+1,+1,-1$)  &($+1,+1,+1,+1,+1,+1,-1$) \\

&S&$D_{2h}^{1}\times Z_{2}^T$&(${\color{red}-1,+1,-1},-1,+1,-1,-1$)  &(${\color{blue}+1,-1,+1},-1,+1,-1,-1$) \\
\hline
\end{tabular}
\end{table*}

\begin{table*}[htbp]
(Continued from the previous page)\\
\centering
\begin{tabular}{ |c|cc|c|c| }
\hline
Shubnikov space group &symmetry point& $G_{\mathbf{Q}}$ &invariants (electron)& invariants (Magnon)\\
\hline

50.284~\textrm{IV}&$\Gamma$&$D_{2h}^{1}\times Z_{2}^T$&($-1,-1,-1,+1,+1,+1,+1$)  &($+1,+1,+1,+1,+1,+1,+1$) \\

&Y&$D_{2h}^{1}\times Z_{2}^T$&($-1,-1,-1,+1,+1,-1,+1$)  &($+1,+1,+1,+1,+1,-1,+1$) \\

&X&$D_{2h}^{1}\times Z_{2}^T$&(${\color{red}-1,+1,-1},-1,+1,+1,-1$)  &(${\color{blue}+1,-1,+1},-1,+1,+1,-1$) \\

&Z&$D_{2h}^{1}\times Z_{2}^T$&($-1,-1,-1,+1,+1,+1,+1$)  &($+1,+1,+1,+1,+1,+1,+1$) \\

&U&$D_{2h}^{1}\times Z_{2}^T$&(${\color{red}-1,+1,-1},-1,+1,+1,-1$)  &(${\color{blue}+1,-1,+1},-1,+1,+1,-1$) \\

&T&$D_{2h}^{1}\times Z_{2}^T$&($-1,-1,-1,+1,+1,-1,+1$)  &($+1,+1,+1,+1,+1,-1,+1$) \\

&S&$D_{2h}^{1}\times Z_{2}^T$&(${\color{red}-1,+1,-1},-1,+1,-1,-1$)  &(${\color{blue}+1,-1,+1},-1,+1,-1,-1$) \\
\hline
\end{tabular}
\end{table*}

\begin{table*}[htbp]
(Continued from the previous page)\\
\centering
\begin{tabular}{ |c|cc|c|c| }
\hline
Shubnikov space group &symmetry point& $G_{\mathbf{Q}}$ &invariants (electron)& invariants (Magnon)\\
\hline

51.301~\textrm{IV}&$\Gamma$&$D_{2h}^{1}\times Z_{2}^T$&($-1,-1,-1,+1,+1,+1,+1$)  &($+1,+1,+1,+1,+1,+1,+1$) \\

&Y&$D_{2h}^{1}\times Z_{2}^T$&(${\color{red}-1,+1,-1},+1,-1,+1,+1$)  &(${\color{blue}+1,-1,+1},+1,-1,+1,+1$) \\

&X&$D_{2h}^{1}\times Z_{2}^T$&($+1,-1,-1,-1,+1,-1,+1$)  &($-1,+1,+1,-1,+1,-1,+1$) \\

&Z&$D_{2h}^{1}\times Z_{2}^T$&(${\color{red}-1,+1,-1},+1,+1,+1,+1$)  &(${\color{blue}+1,-1,+1},+1,+1,+1,+1$) \\

&U&$D_{2h}^{1}\times Z_{2}^T$&(${\color{red}+1,+1,-1},-1,+1,-1,+1$)  &(${\color{blue}-1,-1,+1},-1,+1,-1,+1$) \\

&T&$D_{2h}^{1}\times Z_{2}^T$&($-1,-1,-1,+1,-1,+1,+1$)  &($+1,+1,+1,+1,-1,+1,+1$) \\

&S&$D_{2h}^{1}\times Z_{2}^T$&(${\color{red}+1,+1,-1},-1,-1,-1,+1$)  &(${\color{blue}-1,-1,+1},-1,-1,-1,+1$) \\
\hline
\end{tabular}
\end{table*}

\begin{table*}[htbp]
(Continued from the previous page)\\
\centering
\begin{tabular}{ |c|cc|c|c| }
\hline
Shubnikov space group &symmetry point& $G_{\mathbf{Q}}$ &invariants (electron)& invariants (Magnon)\\
\hline

52.318~\textrm{IV}&$\Gamma$&$D_{2h}^{1}\times Z_{2}^T$&($-1,-1,-1,+1,+1,+1,+1$)  &($+1,+1,+1,+1,+1,+1,+1$) \\

&Y&$D_{2h}^{1}\times Z_{2}^T$&($+1,-1,-1,+1,-1,-1,-1$)  &($-1,+1,+1,+1,-1,-1,-1$) \\

&X&$D_{2h}^{1}\times Z_{2}^T$&(${\color{red}-1,+1,-1},-1,+1,+1,-1$)  &(${\color{blue}+1,-1,+1},-1,+1,+1,-1$) \\

&Z&$D_{2h}^{1}\times Z_{2}^T$&(${\color{red}-1,+1,-1},+1,+1,-1,-1$)  &(${\color{blue}+1,-1,+1},+1,+1,-1,-1$) \\

&U&$D_{2h}^{1}\times Z_{2}^T$&($-1,-1,-1,-1,+1,-1,+1$)  &($+1,+1,+1,-1,+1,-1,+1$) \\

&T&$D_{2h}^{1}\times Z_{2}^T$&(${\color{red}+1,+1,-1},+1,-1,+1,+1$)  &(${\color{blue}-1,-1,+1},+1,-1,+1,+1$) \\

&S&$D_{2h}^{1}\times Z_{2}^T$&(${\color{red}+1,+1,-1},-1,-1,-1,+1$)  &(${\color{blue}-1,-1,+1},-1,-1,-1,+1$) \\
\hline
\end{tabular}
\end{table*}

\begin{table*}[htbp]
(Continued from the previous page)\\
\centering
\begin{tabular}{ |c|cc|c|c| }
\hline
Shubnikov space group &symmetry point& $G_{\mathbf{Q}}$ &invariants (electron)& invariants (Magnon)\\
\hline

53.335~\textrm{IV}&$\Gamma$&$D_{2h}^{1}\times Z_{2}^T$&($-1,-1,-1,+1,+1,+1,+1$)  &($+1,+1,+1,+1,+1,+1,+1$) \\

&Y&$D_{2h}^{1}\times Z_{2}^T$&(${\color{red}-1,+1,-1},+1,-1,+1,+1$)  &(${\color{blue}+1,-1,+1},+1,-1,+1,+1$) \\

&X&$D_{2h}^{1}\times Z_{2}^T$&(${\color{red}-1,+1,-1},-1,+1,+1,-1$)  &(${\color{blue}+1,-1,+1},-1,+1,+1,-1$) \\

&Z&$D_{2h}^{1}\times Z_{2}^T$&($+1,-1,-1,+1,+1,+1,-1$)  &($-1,+1,+1,+1,+1,+1,-1$) \\

&U&$D_{2h}^{1}\times Z_{2}^T$&(${\color{red}+1,+1,-1},-1,+1,+1,+1$)  &(${\color{blue}-1,-1,+1},-1,+1,+1,+1$) \\

&T&$D_{2h}^{1}\times Z_{2}^T$&(${\color{red}+1,+1,-1},+1,-1,+1,-1$)  &(${\color{blue}-1,-1,+1},+1,-1,+1,-1$) \\

&S&$D_{2h}^{1}\times Z_{2}^T$&($-1,-1,-1,-1,-1,+1,-1$)  &($+1,+1,+1,-1,-1,+1,-1$) \\
\hline
\end{tabular}
\end{table*}

\begin{table*}[htbp]
(Continued from the previous page)\\
\centering
\begin{tabular}{ |c|cc|c|c| }
\hline
Shubnikov space group &symmetry point& $G_{\mathbf{Q}}$ &invariants (electron)& invariants (Magnon)\\
\hline

54.352~\textrm{IV}&$\Gamma$&$D_{2h}^{1}\times Z_{2}^T$&($-1,-1,-1,+1,+1,+1,+1$)  &($+1,+1,+1,+1,+1,+1,+1$) \\

&Y&$D_{2h}^{1}\times Z_{2}^T$&(${\color{red}-1,+1,-1},+1,-1,+1,+1$)  &(${\color{blue}+1,-1,+1},+1,-1,+1,+1$) \\

&X&$D_{2h}^{1}\times Z_{2}^T$&(${\color{red}+1,+1,-1},+1,+1,-1,+1$)  &(${\color{blue}-1,-1,+1},+1,+1,-1,+1$) \\

&Z&$D_{2h}^{1}\times Z_{2}^T$&(${\color{red}-1,+1,-1},+1,+1,-1,-1$)  &(${\color{blue}+1,-1,+1},+1,+1,-1,-1$) \\

&U&$D_{2h}^{1}\times Z_{2}^T$&($+1,-1,-1,+1,+1,+1,-1$)  &($-1,+1,+1,+1,+1,+1,-1$) \\

&T&$D_{2h}^{1}\times Z_{2}^T$&($-1,-1,-1,+1,-1,-1,-1$)  &($+1,+1,+1,+1,-1,-1,-1$) \\

&S&$D_{2h}^{1}\times Z_{2}^T$&($+1,-1,-1,+1,-1,-1,+1$)  &($-1,+1,+1,+1,-1,-1,+1$) \\
\hline
\end{tabular}
\end{table*}

\begin{table*}[htbp]
(Continued from the previous page)\\
\centering
\begin{tabular}{ |c|cc|c|c| }
\hline
Shubnikov space group &symmetry point& $G_{\mathbf{Q}}$ &invariants (electron)& invariants (Magnon)\\
\hline

56.374~\textrm{IV}&$\Gamma$&$D_{2h}^{1}\times Z_{2}^T$&($-1,-1,-1,+1,+1,+1,+1$)  &($+1,+1,+1,+1,+1,+1,+1$) \\

&Y&$D_{2h}^{1}\times Z_{2}^T$&(${\color{red}+1,+1,-1},+1,+1,+1,-1$)  &(${\color{blue}-1,-1,+1},+1,+1,+1,-1$) \\

&X&$D_{2h}^{1}\times Z_{2}^T$&($+1,-1,-1,-1,+1,-1,+1$)  &($-1,+1,+1,-1,+1,-1,+1$) \\

&Z&$D_{2h}^{1}\times Z_{2}^T$&(${\color{red}-1,+1,-1},+1,+1,-1,-1$)  &(${\color{blue}+1,-1,+1},+1,+1,-1,-1$) \\

&U&$D_{2h}^{1}\times Z_{2}^T$&(${\color{red}+1,+1,-1},-1,+1,+1,-1$)  &(${\color{blue}-1,-1,+1},-1,+1,+1,-1$) \\

&T&$D_{2h}^{1}\times Z_{2}^T$&($+1,-1,-1,+1,+1,-1,+1$)  &($-1,+1,+1,+1,+1,-1,+1$) \\

&S&$D_{2h}^{1}\times Z_{2}^T$&(${\color{red}-1,+1,-1},-1,+1,-1,-1$)  &(${\color{blue}+1,-1,+1},-1,+1,-1,-1$) \\
\hline
\end{tabular}
\end{table*}

\begin{table*}[htbp]
(Continued from the previous page)\\
\centering
\begin{tabular}{ |c|cc|c|c| }
\hline
Shubnikov space group &symmetry point& $G_{\mathbf{Q}}$ &invariants (electron)& invariants (Magnon)\\
\hline

57.391~\textrm{IV}&$\Gamma$&$D_{2h}^{1}\times Z_{2}^T$&($-1,-1,-1,+1,+1,+1,+1$)  &($+1,+1,+1,+1,+1,+1,+1$) \\

&Y&$D_{2h}^{1}\times Z_{2}^T$&(${\color{red}+1,+1,-1},+1,+1,-1,-1$)  &(${\color{blue}-1,-1,+1},+1,+1,-1,-1$) \\

&X&$D_{2h}^{1}\times Z_{2}^T$&(${\color{red}-1,+1,-1},-1,+1,+1,+1$)  &(${\color{blue}+1,-1,+1},-1,+1,+1,+1$) \\

&Z&$D_{2h}^{1}\times Z_{2}^T$&($+1,-1,-1,+1,+1,+1,-1$)  &($-1,+1,+1,+1,+1,+1,-1$) \\

&U&$D_{2h}^{1}\times Z_{2}^T$&(${\color{red}+1,+1,-1},-1,+1,+1,-1$)  &(${\color{blue}-1,-1,+1},-1,+1,+1,-1$) \\

&T&$D_{2h}^{1}\times Z_{2}^T$&(${\color{red}-1,+1,-1},+1,+1,-1,+1$)  &(${\color{blue}+1,-1,+1},+1,+1,-1,+1$) \\

&S&$D_{2h}^{1}\times Z_{2}^T$&($+1,-1,-1,-1,+1,-1,-1$)  &($-1,+1,+1,-1,+1,-1,-1$) \\
\hline
\end{tabular}
\end{table*}

\begin{table*}[htbp]
(Continued from the previous page)\\
\centering
\begin{tabular}{ |c|cc|c|c| }
\hline
Shubnikov space group &symmetry point& $G_{\mathbf{Q}}$ &invariants (electron)& invariants (Magnon)\\
\hline

59.413~\textrm{IV}&$\Gamma$&$D_{2h}^{1}\times Z_{2}^T$&($-1,-1,-1,+1,+1,+1,+1$)  &($+1,+1,+1,+1,+1,+1,+1$) \\

&Y&$D_{2h}^{1}\times Z_{2}^T$&($+1,-1,-1,+1,-1,+1,-1$)  &($-1,+1,+1,+1,-1,+1,-1$) \\

&X&$D_{2h}^{1}\times Z_{2}^T$&($+1,-1,-1,-1,+1,-1,+1$)  &($-1,+1,+1,-1,+1,-1,+1$) \\

&Z&$D_{2h}^{1}\times Z_{2}^T$&(${\color{red}-1,+1,-1},+1,+1,+1,+1$)  &(${\color{blue}+1,-1,+1},+1,+1,+1,+1$) \\

&U&$D_{2h}^{1}\times Z_{2}^T$&(${\color{red}+1,+1,-1},-1,+1,-1,+1$)  &(${\color{blue}-1,-1,+1},-1,+1,-1,+1$) \\

&T&$D_{2h}^{1}\times Z_{2}^T$&(${\color{red}+1,+1,-1},+1,-1,+1,-1$)  &(${\color{blue}-1,-1,+1},+1,-1,+1,-1$) \\

&S&$D_{2h}^{1}\times Z_{2}^T$&($-1,-1,-1,-1,-1,-1,-1$)  &($+1,+1,+1,-1,-1,-1,-1$) \\
\hline
\end{tabular}
\end{table*}

\begin{table*}[htbp]
(Continued from the previous page)\\
\centering
\begin{tabular}{ |c|cc|c|c| }
\hline
Shubnikov space group &symmetry point& $G_{\mathbf{Q}}$ &invariants (electron)& invariants (Magnon)\\
\hline

60.430~\textrm{IV}&$\Gamma$&$D_{2h}^{1}\times Z_{2}^T$&($-1,-1,-1,+1,+1,+1,+1$)  &($+1,+1,+1,+1,+1,+1,+1$) \\

&Y&$D_{2h}^{1}\times Z_{2}^T$&($-1,-1,-1,+1,+1,-1,+1$)  &($+1,+1,+1,+1,+1,-1,+1$) \\

&X&$D_{2h}^{1}\times Z_{2}^T$&(${\color{red}+1,+1,-1},+1,+1,-1,+1$)  &(${\color{blue}-1,-1,+1},+1,+1,-1,+1$) \\

&Z&$D_{2h}^{1}\times Z_{2}^T$&(${\color{red}+1,+1,-1},+1,+1,+1,-1$)  &(${\color{blue}-1,-1,+1},+1,+1,+1,-1$) \\

&U&$D_{2h}^{1}\times Z_{2}^T$&($-1,-1,-1,+1,+1,-1,-1$)  &($+1,+1,+1,+1,+1,-1,-1$) \\

&T&$D_{2h}^{1}\times Z_{2}^T$&(${\color{red}+1,+1,-1},+1,+1,-1,-1$)  &(${\color{blue}-1,-1,+1},+1,+1,-1,-1$) \\

&S&$D_{2h}^{1}\times Z_{2}^T$&(${\color{red}+1,+1,-1},+1,+1,+1,+1$)  &(${\color{blue}-1,-1,+1},+1,+1,+1,+1$) \\
\hline
\end{tabular}
\end{table*}

\begin{table*}[htbp]
(Continued from the previous page)\\
\centering
\begin{tabular}{ |c|cc|c|c| }
\hline
Shubnikov space group &symmetry point& $G_{\mathbf{Q}}$ &invariants (electron)& invariants (Magnon)\\
\hline

62.452~\textrm{IV}&$\Gamma$&$D_{2h}^{1}\times Z_{2}^T$&($-1,-1,-1,+1,+1,+1,+1$)  &($+1,+1,+1,+1,+1,+1,+1$) \\

&Y&$D_{2h}^{1}\times Z_{2}^T$&($+1,-1,-1,+1,-1,-1,-1$)  &($-1,+1,+1,+1,-1,-1,-1$) \\

&X&$D_{2h}^{1}\times Z_{2}^T$&($+1,-1,-1,-1,+1,-1,+1$)  &($-1,+1,+1,-1,+1,-1,+1$) \\

&Z&$D_{2h}^{1}\times Z_{2}^T$&(${\color{red}+1,+1,-1},+1,+1,-1,+1$)  &(${\color{blue}-1,-1,+1},+1,+1,-1,+1$) \\

&U&$D_{2h}^{1}\times Z_{2}^T$&(${\color{red}-1,+1,-1},-1,+1,+1,+1$)  &(${\color{blue}+1,-1,+1},-1,+1,+1,+1$) \\

&T&$D_{2h}^{1}\times Z_{2}^T$&(${\color{red}-1,+1,-1},+1,-1,+1,-1$)  &(${\color{blue}+1,-1,+1},+1,-1,+1,-1$) \\

&S&$D_{2h}^{1}\times Z_{2}^T$&($-1,-1,-1,-1,-1,+1,-1$)  &($+1,+1,+1,-1,-1,+1,-1$) \\
\hline
\end{tabular}
\end{table*}

\begin{table*}[htbp]
(Continued from the previous page)\\
\centering
\begin{tabular}{ |c|cc|c|c| }
\hline
Shubnikov space group &symmetry point& $G_{\mathbf{Q}}$ &invariants (electron)& invariants (Magnon)\\
\hline

64.479~\textrm{IV}&$\Gamma$&$D_{2h}^{1}\times Z_{2}^T$&($-1,-1,-1,+1,+1,+1,+1$)  &($+1,+1,+1,+1,+1,+1,+1$) \\

&Y&$D_{2h}^{1}\times Z_{2}^T$&($-1,-1,-1,+1,+1,+1,+1$)  &($+1,+1,+1,+1,+1,+1,+1$) \\

&Z&$D_{2h}^{1}\times Z_{2}^T$&($+1,-1,-1,+1,+1,+1,-1$)  &($-1,+1,+1,+1,+1,+1,-1$) \\

&T&$D_{2h}^{1}\times Z_{2}^T$&($+1,-1,-1,+1,+1,+1,-1$)  &($-1,+1,+1,+1,+1,+1,-1$) \\

&S&$C_{2h}^{1}\times Z_{2}^T$&($-1,+1,+1,-1$)  &($-1,-1,+1,+1$) \\

&R&$C_{2h}^{1}\times Z_{2}^T$&($+1,+1,-1,-1$)  &($+1,-1,-1,+1$) \\
\hline
64.480~\textrm{IV}&$\Gamma$&$D_{2h}^{1}\times Z_{2}^T$&($-1,-1,-1,+1,+1,+1,+1$)  &($+1,+1,+1,+1,+1,+1,+1$) \\
\hline
\end{tabular}
\end{table*}

\begin{table*}[htbp]
(Continued from the previous page)\\
\centering
\begin{tabular}{ |c|cc|c|c| }
\hline
Shubnikov space group &symmetry point& $G_{\mathbf{Q}}$ &invariants (electron)& invariants (Magnon)\\
\hline

67.510~\textrm{IV}&$\Gamma$&$D_{2h}^{1}\times Z_{2}^T$&($-1,-1,-1,+1,+1,+1,+1$)  &($+1,+1,+1,+1,+1,+1,+1$) \\

&Y&$D_{2h}^{1}\times Z_{2}^T$&($-1,-1,-1,+1,+1,+1,+1$)  &($+1,+1,+1,+1,+1,+1,+1$)\\

&Z&$D_{2h}^{1}\times Z_{2}^T$&(${\color{red}-1,+1,-1},+1,+1,+1,+1$)  &(${\color{blue}+1,-1,+1},+1,+1,+1,+1$) \\

&T&$D_{2h}^{1}\times Z_{2}^T$&(${\color{red}-1,+1,-1},+1,+1,+1,+1$)  &(${\color{blue}+1,-1,+1},+1,+1,+1,+1$) \\

&S&$C_{2h}^{1}\times Z_{2}^T$&($-1,+1,+1,-1$)  &($-1,-1,+1,+1$) \\

&R&$C_{2h}^{1}\times Z_{2}^T$&($-1,-1,-1,-1$)  &($-1,+1,-1,+1$) \\
\hline

68.518~\textrm{IV}&$\Gamma$&$D_{2h}^{1}\times Z_{2}^T$&($-1,-1,-1,+1,+1,+1,+1$)  &($+1,+1,+1,+1,+1,+1,+1$) \\
\hline
\end{tabular}
\end{table*}

\begin{table*}[htbp]
(Continued from the previous page)\\
\centering
\begin{tabular}{ |c|cc|c|c| }
\hline
Shubnikov space group &symmetry point& $G_{\mathbf{Q}}$ &invariants (electron)& invariants (Magnon)\\
\hline

74.562~\textrm{IV}&$\Gamma$&$D_{2h}^{1}\times Z_{2}^T$&($-1,-1,-1,+1,+1,+1,+1$)  &($+1,+1,+1,+1,+1,+1,+1$) \\

&X&$D_{2h}^{1}\times Z_{2}^T$&($-1,-1,-1,+1,+1,+1,+1$)  &($+1,+1,+1,+1,+1,+1,+1$) \\

&R&$C_{2h}^{3}\times Z_{2}^T$&($+1,+1,+1,-1$)  &($+1,-1,+1,+1$) \\

&S&$C_{2h}^{2}\times Z_{2}^T$&($+1,-1,+1,-1$)  &($+1,+1,+1,+1$) \\

&T&$C_{2h}^{1}\times Z_{2}^T$&($-1,+1,+1,-1$)  &($-1,-1,+1,+1$) \\

&W&$D_{2}\times Z_{2}^T$&($+1,-1,+1,-1$)  &($-1,+1,+1,-1$) \\
\hline
83.48~\textrm{IV}&$\Gamma$&$C_{4h}\times Z_{2}^T$&($+1,-1,-1,+1$) &($+1,+1,+1,+1$) \\
\hline
\end{tabular}
\end{table*}

\clearpage
\begin{table*}[htbp]
(Continued from the previous page)\\
\centering
\begin{tabular}{ |c|cc|c|c| }
\hline
Shubnikov space group &symmetry point& $G_{\mathbf{Q}}$ &invariants (electron)& invariants (Magnon)\\
\hline

86.72~\textrm{IV}&$\Gamma$&$C_{4h}\times Z_{2}^T$&($+1,-1,-1,+1$) &($+1,+1,+1,+1$) \\

&M&$C_{4h}\times Z_{2}^T$&($-1,-1,-1,+1$) &($-1,+1,+1,+1$) \\

&Z&$C_{4h}\times Z_{2}^T$&($-1,+1,-1,+1$) &($-1,-1,+1,+1$) \\

&A&$C_{4h}\times Z_{2}^T$&($+1,+1,-1,+1$) &($+1,-1,+1,+1$) \\

&R&$C_{2h}^{1}\times Z_{2}^T$&($-1,+1,-1,-1$) &($-1,-1,-1,+1$) \\

&X&$C_{2h}^{1}\times Z_{2}^T$&($-1,-1,+1,-1$) &($-1,+1,+1,+1$) \\
\hline

86.73~\textrm{IV}&$\Gamma$&$C_{4h}\times Z_{2}^T$&($+1,-1,-1,+1$) &($+1,+1,+1,+1$) \\
\hline
\end{tabular}
\end{table*}

\begin{table*}[htbp]
(Continued from the previous page)\\
\centering
\begin{tabular}{ |c|cc|c|c| }
\hline
Shubnikov space group &symmetry point& $G_{\mathbf{Q}}$ &invariants (electron)& invariants (Magnon)\\
\hline

125.372~\textrm{IV}&$\Gamma$&$D_{4h}^{1}\times Z_{2}^T$&($-1,-1,-1,+1,+1,+1,+1$) &($+1,+1,+1,+1,+1,+1,+1$) \\

&M&$D_{4h}^{1}\times Z_{2}^T$&($-1,-1,-1,+1,+1,-1,+1$) &($+1,+1,+1,+1,+1,-1,+1$) \\

&Z&$D_{4h}^{1}\times Z_{2}^T$&(${\color{red}-1,+1,-1},+1,+1,+1,+1$) &(${\color{blue}+1,-1,+1},+1,+1,+1,+1$) \\

&A&$D_{4h}^{1}\times Z_{2}^T$&(${\color{red}-1,+1,-1},+1,+1,-1,+1$) &(${\color{blue}+1,-1,+1},+1,+1,-1,+1$) \\

&R&$D_{2h}^{1}\times Z_{2}^T$&(${\color{red}-1,+1,-1},+1,+1,-1,+1$)  &(${\color{blue}+1,-1,+1},+1,+1,-1,+1$) \\

&X&$D_{2h}^{1}\times Z_{2}^T$&($-1,-1,-1,+1,+1,-1,+1$)  &($+1,+1,+1,+1,+1,-1,+1$) \\
\hline
125.373~\textrm{IV}&$\Gamma$&$D_{4h}^{1}\times Z_{2}^T$&($-1,-1,-1,+1,+1,+1,+1$) &($+1,+1,+1,+1,+1,+1,+1$) \\
\hline
\end{tabular}
\end{table*}

\begin{table*}[htbp]
(Continued from the previous page)\\
\centering
\begin{tabular}{ |c|cc|c|c| }
\hline
Shubnikov space group &symmetry point& $G_{\mathbf{Q}}$ &invariants (electron)& invariants (Magnon)\\
\hline

128.409~\textrm{IV}&$\Gamma$&$D_{4h}^{1}\times Z_{2}^T$&($-1,-1,-1,+1,+1,+1,+1$) &($+1,+1,+1,+1,+1,+1,+1$) \\

&M&$D_{4h}^{1}\times Z_{2}^T$&($-1,-1,-1,+1,+1,+1,+1$) &($+1,+1,+1,+1,+1,+1,+1$)\\

&Z&$D_{4h}^{1}\times Z_{2}^T$&($-1,-1,-1,+1,+1,-1,-1$) &($+1,+1,+1,+1,+1,-1,-1$) \\

&A&$D_{4h}^{1}\times Z_{2}^T$&($-1,-1,-1,+1,+1,-1,-1$) &($+1,+1,+1,+1,+1,-1,-1$) \\

&R&$D_{2h}^{1}\times Z_{2}^T$&(${\color{red}+1,+1,-1},+1,+1,+1,+1$)  &(${\color{blue}-1,-1,+1},+1,+1,+1,+1$) \\

&X&$D_{2h}^{1}\times Z_{2}^T$&(${\color{red}+1,+1,-1},+1,+1,-1,-1$)  &(${\color{blue}-1,-1,+1},+1,+1,-1,-1$) \\
\hline

128.410~\textrm{IV}&$\Gamma$&$D_{4h}^{1}\times Z_{2}^T$&($-1,-1,-1,+1,+1,+1,+1$) &($+1,+1,+1,+1,+1,+1,+1$) \\
\hline
\end{tabular}
\end{table*}

\begin{table*}[htbp]
(Continued from the previous page)\\
\centering
\begin{tabular}{ |c|cc|c|c| }
\hline
Shubnikov space group &symmetry point& $G_{\mathbf{Q}}$ &invariants (electron)& invariants (Magnon)\\
\hline

131.446~\textrm{IV}&$\Gamma$&$D_{4h}^{1}\times Z_{2}^T$&($-1,-1,-1,+1,+1,+1,+1$) &($+1,+1,+1,+1,+1,+1,+1$) \\

&M&$D_{4h}^{1}\times Z_{2}^T$&($-1,-1,-1,-1,+1,+1,+1$) &($+1,+1,+1,-1,+1,+1,+1$) \\

&Z&$D_{4h}^{1}\times Z_{2}^T$&(${\color{red}-1,+1,-1},+1,+1,+1,-1$) &(${\color{blue}+1,-1,+1},+1,+1,+1,-1$) \\

&A&$D_{4h}^{1}\times Z_{2}^T$&(${\color{red}-1,+1,-1},-1,+1,+1,-1$) &(${\color{blue}+1,-1,+1},-1,+1,+1,-1$) \\

&R&$D_{2h}^{1}\times Z_{2}^T$&($-1,-1,-1,+1,-1,+1,+1$)  &($+1,+1,+1,+1,-1,+1,+1$) \\

&X&$D_{2h}^{1}\times Z_{2}^T$&(${\color{red}-1,+1,-1},+1,-1,+1,+1$)  &(${\color{blue}+1,-1,+1},+1,-1,+1,+1$) \\
\hline

132.456~\textrm{IV}&$\Gamma$&$D_{4h}^{1}\times Z_{2}^T$&($-1,-1,-1,+1,+1,+1,+1$) &($+1,+1,+1,+1,+1,+1,+1$) \\
\hline
\end{tabular}
\end{table*}

\begin{table*}[htbp]
(Continued from the previous page)\\
\centering
\begin{tabular}{ |c|cc|c|c| }
\hline
Shubnikov space group &symmetry point& $G_{\mathbf{Q}}$ &invariants (electron)& invariants (Magnon)\\
\hline

135.492~\textrm{IV}&$\Gamma$&$D_{4h}^{1}\times Z_{2}^T$&($-1,-1,-1,+1,+1,+1,+1$) &($+1,+1,+1,+1,+1,+1,+1$) \\

&M&$D_{4h}^{1}\times Z_{2}^T$&($-1,-1,-1,-1,+1,+1,+1$) &($+1,+1,+1,-1,+1,+1,+1$) \\

&Z&$D_{4h}^{1}\times Z_{2}^T$&(${\color{red}-1,+1,-1},+1,+1,+1,-1$) &(${\color{blue}+1,-1,+1},+1,+1,+1,-1$) \\

&A&$D_{4h}^{1}\times Z_{2}^T$&(${\color{red}-1,+1,-1},-1,+1,+1,-1$) &(${\color{blue}+1,-1,+1},-1,+1,+1,-1$) \\

&R&$D_{2h}^{1}\times Z_{2}^T$&(${\color{red}+1,+1,-1},+1,-1,-1,-1$)  &(${\color{blue}-1,-1,+1},+1,-1,-1,-1$) \\

&X&$D_{2h}^{1}\times Z_{2}^T$&($+1,-1,-1,+1,-1,-1,-1$)  &($-1,+1,+1,+1,-1,-1,-1$) \\
\hline
135.493~\textrm{IV}&$\Gamma$&$D_{4h}^{1}\times Z_{2}^T$&($-1,-1,-1,+1,+1,+1,+1$) &($+1,+1,+1,+1,+1,+1,+1$) \\
\hline
\end{tabular}
\end{table*}

\begin{table*}[htbp]
(Continued from the previous page)\\
\centering
\begin{tabular}{ |c|cc|c|c| }
\hline
Shubnikov space group &symmetry point& $G_{\mathbf{Q}}$ &invariants (electron)& invariants (Magnon)\\
\hline

138.529~\textrm{IV}&$\Gamma$&$D_{4h}^{1}\times Z_{2}^T$&($-1,-1,-1,+1,+1,+1,+1$) &($+1,+1,+1,+1,+1,+1,+1$) \\

&M&$D_{4h}^{1}\times Z_{2}^T$&($-1,-1,-1,+1,+1,-1,+1$) &($+1,+1,+1,+1,+1,-1,+1$) \\

&Z&$D_{4h}^{1}\times Z_{2}^T$&($-1,-1,-1,+1,+1,-1,+1$) &($+1,+1,+1,+1,+1,-1,+1$) \\

&A&$D_{4h}^{1}\times Z_{2}^T$&($-1,-1,-1,+1,+1,+1,+1$) &($+1,+1,+1,+1,+1,+1,+1$)\\

&R&$D_{2h}^{1}\times Z_{2}^T$&(${\color{red}+1,+1,-1},+1,+1,-1,+1$)  &(${\color{blue}-1,-1,+1},+1,+1,-1,+1$) \\

&X&$D_{2h}^{1}\times Z_{2}^T$&(${\color{red}+1,+1,-1},+1,+1,+1,-1$)  &(${\color{blue}-1,-1,+1},+1,+1,+1,-1$) \\
\hline
138.530~\textrm{IV}&$\Gamma$&$D_{4h}^{1}\times Z_{2}^T$&($-1,-1,-1,+1,+1,+1,+1$) &($+1,+1,+1,+1,+1,+1,+1$) \\
\hline
\end{tabular}
\end{table*}

\begin{table*}[htbp]
(Continued from the previous page)\\
\centering
\begin{tabular}{ |c|cc|c|c| }
\hline
Shubnikov space group &symmetry point& $G_{\mathbf{Q}}$ &invariants (electron)& invariants (Magnon)\\
\hline

165.96~\textrm{IV}&$\Gamma$&$D_{3d}^{2}\times Z_{2}^T$&($+1,-1,-1,+1$) &($+1,+1,+1,+1$) \\

&M&$C_{2h}^{7}\times Z_{2}^T$&($+1,-1,+1,-1$) &($+1,+1,+1,+1$) \\

&A&$D_{3d}^{2}\times Z_{2}^T$&($-1,+1,-1,+1$) &($-1,-1,+1,+1$)\\

&L&$C_{2h}^{7}\times Z_{2}^T$&($-1,+1,+1,-1$) &($-1,-1,+1,+1$) \\

&K&$D_{3}^{}\times Z_{2}^T$&($-1,+1$) &($+1,+1$) \\

&H&$D_{3}^{}\times Z_{2}^T$&($-1,+1$) &($+1,+1$) \\
\hline

166.102~\textrm{IV}&$\Gamma$&$D_{3d}^{1}\times Z_{2}^T$&($+1,-1,-1,+1$) &($+1,+1,+1,+1$) \\
\hline
\end{tabular}
\end{table*}

\begin{table*}[htbp]
(Continued from the previous page)\\
\centering
\begin{tabular}{ |c|cc|c|c| }
\hline
Shubnikov space group &symmetry point& $G_{\mathbf{Q}}$ &invariants (electron)& invariants (Magnon)\\
\hline

203.29~\textrm{IV}&$\Gamma$&$T_{h}\times Z_{2}^T$&($-1,-1,-1$) &($+1,+1,+1$) \\

&X&$D_{2h}^{1}\times Z_{2}^T$&($-1,-1,-1,+1,+1,-1,+1$)  &($+1,+1,+1,+1,+1,-1,+1$) \\

&L&$C_{3i}^{2}\times Z_{2}^T$&($+1,-1$) &($-1,+1$) \\

&W&$C_{2v}\times Z_{2}^T$&($+1,-1,-1,+1$) &($-1,+1,-1,+1$) \\
\hline

205.36~\textrm{IV}&$\Gamma$&$T_{h}\times Z_{2}^T$&($-1,-1,-1$) &($+1,+1,+1$) \\

&X&$D_{2h}^{1}\times Z_{2}^T$&(${\color{red}+1,+1,-1},+1,+1,-1,-1$)  &(${\color{blue}-1,-1,+1},+1,+1,-1,-1$) \\

&M&$D_{2h}^{1}\times Z_{2}^T$&($-1,-1,-1,+1,+1,+1,-1$)  &($+1,+1,+1,+1,+1,+1,-1$) \\
\hline
\end{tabular}
\end{table*}

\end{document}